\documentclass[aps,11pt,prd,groupedaddress,nofootinbib,notitlepage,eqsecnum,preprintnumbers]{revtex4-2}
\usepackage[utf8]{inputenc}
\usepackage{hyperref}
\usepackage{xcolor}
\usepackage{graphicx}
\usepackage{amsmath,amssymb}
\usepackage{bm}
\usepackage{comment}
\usepackage[shortlabels]{enumitem}
\usepackage{overpic}
\usepackage{ulem}

\def\keq{k_{\rm eq}}

\begin{document}
\title{Cosmological Gravitational Waves from Isocurvature Fluctuations}

\author{\textsc{Guillem Dom\`enech$^{a,b}$}}
    \email{{guillem.domenech}@{itp.uni-hannover.de}}

\affiliation{$^a$ Institut für Theoretische Physik, Leibniz Universität Hannover, Appelstraße 2, 30167 Hannover, Germany.}
\affiliation{$^b$ Max-Planck-Institut für Gravitationsphysik, Albert-Einstein-Institut, 30167 Hannover, Germany}

\begin{abstract}
Gravitational waves induced by large primordial curvature fluctuations may result in a sizable stochastic gravitational wave background. Interestingly, curvature fluctuations are gradually generated by initial isocurvature fluctuations, which in turn induce gravitational waves. Initial isocurvature fluctuations commonly appear in multi-field models of inflation as well as in the formation of scattered compact objects in the very early universe, such as primordial black holes and solitons like oscillons and cosmic strings. Here we provide a  review on isocurvature induced gravitational waves and its applications to dark matter and the primordial black hole dominated early universe.
\end{abstract}

\maketitle

\section{Introduction \label{sec:intro}}

The standard model of cosmology requires adiabatic primordial fluctuations as initial conditions set in the very early universe \cite{WMAP:2003ivt,Akrami:2018odb,dodelson2021modern}. Fluctuations are adiabatic when all fields filling the universe agreed to share a slicing of the spacetime where each of its separate energy density fluctuations vanish. In such choice of coordinates the energy density of those fields is homogeneous (by definition) and only the metric contains fluctuations. These are the so-called adiabatic curvature fluctuations. The reason for this sort of coherent initial conditions may simply be that one “primordial” field is responsible for all the fluctuations, e.g. because it decays into all the other fields. This is precisely the prediction from the simplest models of cosmic inflation and what Cosmic Microwave Background (CMB) observations confirmed \cite{Planck:2018vyg}.

If the initial conditions for primordial fluctuations are not adiabatic, they are said to be isocurvature (see e.g. \cite{Kodama:1985bj,Kodama:1986fg,Kodama:1986ud,Langlois:2003fq,Bucher:1999re}). As the name implies, isocurvature requires that there is no initial adiabatic curvature perturbation. This means that there is no slicing of spacetime where each of the energy density fluctuations separately vanish (although there is always the slicing where the total energy density fluctuation is zero). And so, isocurvature initial conditions are related to relative energy density fluctuations (to be precise relative number density fluctuations). On scales larger than $10\,{\rm Mpc}$, CMB tells us that isocurvature primordial fluctuations may not account for more than $1-10\%$ of the total fluctuations \cite{Akrami:2018odb}. Since the measured amplitude of the power spectrum of primordial adiabatic fluctuations is about $10^{-9}$, the power spectrum of isocurvature fluctuations may have an amplitude of less than $10^{-10}$ on large scales.

The story changes when we consider scales smaller than $1\,\,{\rm Mpc}$, where CMB constrains no longer apply. For scales between $1\,{\rm Mpc}$ and $1\,{\rm pc}$ future CMB spectral distortions might be able to test isocurvature fluctuations \cite{Chluba:2013dna,Chluba:2019kpb}.\footnote{Baryon isocurvature is constrained by Big Bang Nucleosynthesis \cite{Inomata:2018htm} to be less than $1\%$ in a similar range of scales.} For scales smaller than a parsec our best bet to constrain the amplitude and nature of primordial fluctuations are Primordial Black Holes (PBHs) and induced Gravitational Waves (GWs).\footnote{On those scales, one may constrain CDM fluctuations if are non-trivial particle interactions with CDM \cite{Kohri:2014lza,Yang:2014lsg,Nakama:2017qac}.} PBHs form from the collapse of large primordial fluctuations \cite{Zeldovich:1967lct,Hawking:1971ei,Carr:1974nx,Meszaros:1974tb,Carr:1975qj,Khlopov:1985jw,Niemeyer:1999ak} and could explain the Cold Dark Matter (CDM) \cite{Bellomo:2017zsr,Carr:2017jsz,Inomata:2017okj,Bartolo:2018rku,Bartolo:2018evs,Carr:2020xqk}, some of the LIGO/VIRGO/KAGRA black hole mergers \cite{Bird:2016dcv,Sasaki:2016jop,Wong:2020yig,Franciolini:2021tla} and be the seeds of supermassive black holes \cite{Kawasaki:2012kn,Carr:2018rid}. Induced GWs are a consequence of the non-linear nature of gravity: density fluctuations eventually lead to metric fluctuations, which include GWs as a secondary effect \cite{Tomita,Matarrese:1992rp,Matarrese:1993zf,Ananda:2006af,Baumann:2007zm,Saito:2008jc,Saito:2009jt}. In some sense, the evolution of primordial fluctuations (e.g. the resulting density waves) yield an anisotropic stress which sources the secondary, or induced, GWs. See Refs.~\cite{Khlopov:2008qy,Sasaki:2018dmp,Carr:2020gox,Green:2020jor,Escriva:2022duf,Ozsoy:2023ryl} and \cite{Domenech:2021ztg,Yuan:2021qgz} for recent reviews on PBHs and induced GWs respectively. See also Ref.~\cite{Domenech:2023fuz} for lecture notes on the collection of GW signatures of PBHs.

By extension of the CMB, the standard approach is to consider adiabatic initial conditions for PBHs and induced GWs. But, both PBHs and induced GWs may also be generated from isocurvature initial conditions \cite{Passaglia:2021jla,Domenech:2021and}.\footnote{Other PBH formation scenarios include first-order phase transitions \cite{Crawford:1982yz,Kodama:1982sf}, collapse of Q-balls \cite{Cotner:2016cvr,Cotner:2019ykd,Flores:2021jas} and long-range forces in the early universe \cite{Amendola:2017xhl,Flores:2020drq,Domenech:2023afs}.} Isocurvature fluctuations generally occur in multi-field models of inflation \cite{Chung:2017uzc,Chung:2021lfg}, phase transitions \cite{Dolgov:1992pu} and may be consequence of the Poisson noise (or clustering) from the formation of compact structures in the very early universe, such as PBHs and solitonic structures like oscillons and cosmic strings \cite{Inman:2019wvr,Cotner:2019ykd,Papanikolaou:2020qtd,Lozanov:2023aez,Lozanov:2023knf}. This opens the door to probes of the nature of primordial fluctuations and the formation of compact objects in the early universe. In this review we will focus on isocurvature induced GWs. Details on PBHs formed by the collapse of primordial CDM isocurvature fluctuations can be found in Ref.~\cite{Passaglia:2021jla}.

We are in an exciting time for cosmology with GWs. Recently, Pulsar Timing Array (PTA) collaborations around the globe have announced tentative evidence of a GW background at nHz frequencies \cite{NG15-SGWB,NG15-pulsars,EPTA2-SGWB,EPTA2-pulsars,EPTA2-SMBHB-NP,PPTA3-SGWB,PPTA3-pulsars,PPTA3-SMBHB,CPTA-SGWB,InternationalPulsarTimingArray:2023mzf}. On plausible explanation is that they are induced GWs from primordial fluctuations \cite{NG15-NP,Dandoy:2023jot,Franciolini:2023pbf,Franciolini:2023wjm,Inomata:2023zup,Cai:2023dls,Wang:2023ost,Liu:2023ymk,Unal:2023srk,Figueroa:2023zhu,Yi:2023mbm,Zhu:2023faa,Firouzjahi:2023lzg,Li:2023qua,You:2023rmn,Balaji:2023ehk,HosseiniMansoori:2023mqh,Zhao:2023joc,Liu:2023pau,Yi:2023tdk,Bhaumik:2023wmw,Choudhury:2023hfm,Yi:2023npi,Harigaya:2023pmw,Basilakos:2023xof,Jin:2023wri,Cannizzaro:2023mgc,Zhang:2023nrs,Liu:2023hpw,Choudhury:2023fwk,Tagliazucchi:2023dai,Basilakos:2023jvp,Inomata:2023drn,Li:2023xtl,Domenech:2023dxx,Gangopadhyay:2023qjr,Cyr:2023pgw} (or the merger of supermassive PBHs \cite{Huang:2023chx,Gouttenoire:2023nzr,Depta:2023qst}). As a new application, here we will also explore the possibility that the PTA results are explained by CDM isocurvature induced GWs. The PTA results are and will be complemented by other GW detectors at higher frequencies such as the LIGO/Virgo/KAGRA collaboration \cite{KAGRA:2021kbb} and future GW detectors like $\mu$-Ares \cite{Sesana:2019vho}, LISA, Taiji \cite{Barke:2014lsa,Ruan:2018tsw}, TianQin \cite{Gong:2021gvw},  DECIGO \cite{Yagi:2011wg,Kawamura:2020pcg}, Einstein Telescope (ET) \cite{Maggiore:2019uih}, Cosmic Explorer (CE) \cite{ce}, Voyager \cite{A+,voyager}.

This work is organized as follows. In \S~\ref{sec:review} we present an overview of the basic equations and the notion of adiabatic and isocurvature initial conditions. The content of this section is based on Refs.~\cite{Kodama:1986ud,Langlois:2003fq}. Then, we discuss some applications. In \S~\ref{sec:GWCDM} we view GWs induced by CDM, based on the results of Ref.~\cite{Domenech:2021and}. In \S~\ref{sec:PBHdom} we consider the isocurvature due to PBH number density fluctuations in the PBH reheating scenario. Here we will base our discussions on Refs.~\cite{Inomata:2020lmk,Papanikolaou:2020qtd,Domenech:2020ssp,Domenech:2021wkk}. Lastly, we discuss more applications in \S~\ref{sec:conclusions}. Some details of the calculations can be found in the appendices and in the aforementioned references. We work in reduced Planck units where $c=\hbar=1$ and $M_{\rm pl}=(8\pi G)^{-1/2}=1$.

\section{Basic equations \label{sec:review}}

We start by deriving the general formalism for isocurvature induced GWs. Since isocurvature fluctuations result from relative number density fluctuations, we should at least consider two fluids filling the primordial universe. For simplicity, we assume that after cosmic inflation the universe is dominated by relativistic particles, so-called radiation, and that there is a small but non-zero fraction of non-relativistic particles, let us call it matter. The energy momentum tensors of radiation and matter are respectively given by
\begin{align}\label{eq:tmunus}
T_{r\mu\nu}&=(\rho_r+p_r)u_{r\mu}u_{r\nu}+p_r g_{\mu\nu}\,,\\
T_{m\mu\nu}&=\rho_m u_{m\mu}u_{m\nu}\,,
\end{align}
where $g_{\mu\nu}$ is the metric, $\rho$ and $p$ respectively are the energy density and pressure and $u_{\mu}$ the fluid $4$-velocity. The subscripts ``$r$'' and ``$m$'' respectively refer to radiation and matter. In particular, we have that $p_m=0$ and $p_r=\rho_r/3$. For the metric $g_{\mu\nu}$ we take a perturbed Friedmann–Lemaître–Robertson–Walker (FLRW) universe. For convenience (the equations for induced GWs are simplest), we work in the Newtonian gauge in which the metric reads
\begin{align}
ds^2=a^2(\tau)\left[-(1+2\Psi)d\tau^2+((1+2\Phi)\delta_{ij}+h_{ij})dx^idx^j\right]\,,
\end{align}
where $a$ is the scale factor, $\tau$ is conformal time, $\Psi$ and $\Phi$ are the gravitational potentials and $h_{ij}$ the tensor perturbations (or we may say GWs). The dynamics of the scale factor are dictated by the Friedmann equations, which we present in App.~\ref{app:cosmoformulas}. In the same appendix we describe the notation for matter perturbations as well. 

Most important to us is the fact that energy conservation ($\nabla^\mu T_{\mu\nu}=0$) at the background level requires that $\rho_r\propto a^{-4}$ and $\rho_m\propto a^{-3}$ (also see App.~\ref{app:cosmoformulas}). From the different dilution of the energy densities, we see that
\begin{align}
\frac{\rho_m}{\rho_r}\propto a\,,
\end{align}
and, therefore, matter fields eventually dominate the energy density of the universe, assuming they do not decay. Thus, if we call $\beta$ to the initial fraction of matter, namely 
\begin{align}
\beta=\frac{\rho_{m,i}}{\rho_{r,i}}<1\,,
\end{align}
where “$i$” refers to the initial time, the universe will be matter dominated after $a/a_i>\beta^{-1}$. Actually, there is an exact analytical solution for the scale factor in the radiation-matter universe. It reads \cite{Kodama:1986ud,mukhanov2005physical}
\begin{align}\label{eq:scalefactor}
\frac{a(\tau)}{a_{\rm eq}}=2\left(\frac{\tau}{\tau_*}\right)+\left(\frac{\tau}{\tau_*}\right)^2\,,
\end{align}
where $(\sqrt{2}-1)\tau_*=\tau_{\rm eq}$. The subscript ``eq'' means matter-radiation equality, i.e. the time when $\rho_r=\rho_m$. It is easy to check that Eq.~\eqref{eq:scalefactor} goes from the radiation dominated universe where $a\sim \tau$ to the matter dominated universe with $a\sim \tau^2$. This is for the moment all we need to understand the source of induced gravitational waves.

\subsection{What sources secondary gravitational waves?}

Before entering into the computational details, let us qualitatively understand what is the main source of secondary gravitational waves (at least in the Newton gauge). Let us formally start with Einstein equations, that is
\begin{align}
G_{\mu\nu}=T_{m\mu\nu}+T_{r\mu\nu}\,,
\end{align}
where $G_{\mu\nu}$ is the Einstein tensor and for simplicity we set $M_{\rm pl}=(8\pi G)^{-1/2}=1$. The linear equations of motion for the tensor modes correspond to the transverse-traceless projection of the spatial-spatial components. If we call ${\cal P}_{ij}\,^{ab}$ the transverse-traceless projector (which can be found in, e.g., Refs.~\cite{Domenech:2020ssp,Domenech:2021ztg}), then we schematically have at linear order
\begin{align}\label{eq:linearGmunu}
{\cal P}_{ij}\,^{ab}G^{(1)}_{ab}={\cal P}_{ij}\,^{ab}\left(T^{(1)}_{mab}+T^{(1)}_{rab}\right)\quad \Rightarrow\quad h_{ij}''+2{\cal H}h'_{ij}+\Delta h_{ij}=0\,,
\end{align}
where ${\cal H}=a'/a$, $'=d/d\tau$ and the superscript $(1)$ refers to linear perturbations. We will use a superscript $(2)$ to refer to second order perturbations. Eq.~\eqref{eq:linearGmunu} basically tells us that since $T_{\mu\nu}$ has no tensor component at linear order (because we assumed a perfect fluid with no anisotropic stress), gravitational waves propagate freely. If we include second order scalar terms though, we find that
\begin{align}\label{eq:linearGmunu2}
{\cal P}_{ij}\,^{ab}G^{(1)}_{ab}={\cal P}_{ij}\,^{ab}\left(-G^{(2)}_{ab}+T^{(2)}_{mab}+T^{(2)}_{rab}\right)\quad \Rightarrow\quad h_{ij}''+2{\cal H}h'_{ij}+\Delta h_{ij}={\cal P}_{ij}\,^{ab}S_{ab}\,,
\end{align}
which after some simplifications the source term is given by \cite{Domenech:2020ssp}
\begin{align}\label{eq:source}
S_{ij}&=4\partial_i\Phi\partial_j\Phi+2a^2\rho_m \partial_iv_{m}\partial_jv_{m}+2a^2(\rho_r+p_r)\partial_iv_{r}\partial_iv_{r}\,.
\end{align}
Note that we selected the scalar component of the perturbation of the spatial velocity in Eq.~\eqref{eq:tmunus}, that is we took $u_i=a\partial_iv$.
Also in Eq.~\eqref{eq:linearGmunu2} we considered the second order expansion of $G^{(2)}_{ab}$ as a source (or backreaction) to the linear equations and, as such, we moved it to the right hand side. We then used that $\Psi+\Phi=0$ in the presence of no anisotropic stress (see App.~\ref{app:cosmoformulas}).

Let us discuss the secondary source to GWs, Eq.~\eqref{eq:source}, within the big picture. First, we see that from $G^{(2)}_{ab}$ we obtain gradients of $\Phi$. So, one source of secondary gravitational waves are curvature (metric) fluctuations. Second, from $T_{\mu\nu}$ only the spatial component of the fluid velocity contributes. So velocity flows also generate GWs (this is the main, intuitive, source of GWs inside the cosmic horizon). However, this does not tells us much yet about isocurvature initial conditions. To make it more intuitive, let us introduce the total spatial velocity (the one seen by the linear Einstein Equations and so linear metric fluctuations) and the relative velocity which respectively read \cite{Malik_2009}
\begin{align}
\rho V=(\rho_m+\rho_r) V=\rho_m v_m+(\rho_r+p_r) v_r\quad{\rm and}\quad V_{\rm rel}=v_m-v_r \,.
\end{align}
In terms of these variables we find that Eq.~\eqref{eq:source} is given by
\begin{align}\label{eq:source2}
S_{ij}&=4\partial_i\Phi\partial_j\Phi+2a^2\frac{\rho^2}{\rho+p}\left(\partial_i V\partial_j V+\frac{\rho_m(\rho_r+p_r)}{\rho^2}\partial_i V_{\rm rel}\partial_j V_{\rm rel}\right)\,,
\end{align}
where $p=p_m+p_r=p_r$.
Now, it is clear that the contribution from the relative velocity is always suppressed by the energy density of the subdominant field. So unless we are considering scales that enter the horizon close to the matter-radiation equality, we may neglect the relative velocities. With this we conclude that secondary gravitational waves are mainly sourced by the dominant fluid, which is the main source of curvature fluctuations. Any contribution from isocurvature initial conditions (= no initial curvature or $\Phi_i=0$) must then be suppressed by a factor $\rho_m/\rho$ in the early stages.

A natural question then arises: why should we consider isocurvature induced GWs? But this question misses that the point that isocurvature is a matter of initial conditions. As the system evolves, isocurvature fluctuations are transferred into curvature fluctuations, the transfer being complete after matter-radiation equality \cite{Kodama:1986ud}. So the answer is that isocurvature induced GWs can be important when:
\begin{itemize}
\item isocurvature fluctuations are large enough to compensate for the suppression factor $\rho_m/\rho$, or,
\item non-relativistic particles dominate the universe in an early matter dominated era with isocurvature induced curvature fluctuations.
\end{itemize}
We will study the first case in \S~\ref{sec:GWCDM} for CDM fluctuations and the second in \S~\ref{sec:PBHdom} for the PBH dominated early universe. We now proceed with the evolution of the curvature fluctuations and the formal solutions to induced GWs.

\subsection{Evolution of curvature perturbation: adiabatic vs isocurvature initial conditions}

The general notion of curvature and isocurvature fluctuations is better understood using the gauge invariant definition of the curvature perturbation on uniform density slices, which is given by, see e.g. \cite{Langlois:2003fq,Lyth:2004gb},
\begin{align}\label{eq:defzeta}
\zeta=-\phi+{\cal H}\frac{\delta\rho}{\rho'}\,.
\end{align}
Here we used $\phi$ to denote the spatial curvature perturbation which corresponds to $\Phi$ in the Newton gauge. In the uniform density slicing where $\delta\rho=0$ we have that $\zeta=-\phi$. The convenience of using $\zeta$ is that it is conserved on superhorizon scales. One may also define a curvature fluctuation for each fluid, namely
\begin{align}
\zeta_y=-\phi+{\cal H}\frac{\delta\rho_y}{\rho'_y}\,,
\end{align}
where in our case $y=\{r,m\}$. And, from the individual definitions, the notion of isocurvature follows as the relative individual curvature fluctuations, namely
\begin{align}\label{eq:defiso}
S=3(\zeta_r-\zeta_m)=\frac{\delta\rho_m}{\rho_m}-\frac{\delta\rho_r}{\rho_r+p_r}=\frac{\delta\rho_m}{\rho_m}-\frac{3}{4}\frac{\delta\rho_r}{\rho_r}\,,
\end{align}
where we used that $\rho'_y+3{\cal H}(\rho_y+p_y)=0$ and that $p_r=1/3 \rho_r$. The physical intuition behind such formal definitions, Eqs.~\eqref{eq:defzeta} and \eqref{eq:defiso}, is the following. Curvature fluctuations are metric fluctuations even in the time slicing where there are no total density fluctuations ($\delta\rho=0$). Adiabatic initial conditions then require no initial isocurvature, that is $S_i=0$, and $\zeta_i\neq 0$. Isocurvature initial conditions correspond to the case when there are no such curvature fluctuations but there are relative density fluctuations (i.e. we may have $\zeta_i=0$ but $S_i\neq 0$). 

To understand this also in terms of the Newton gauge variables, it is convenient to look at the $0$-$0$ Einstein Equation on superhorizon scales (see App.~\ref{app:cosmoformulas}), which reads
\begin{align}
6{\cal H}_i^2\Phi_i=a_i^2\left(\delta\rho_{m,i}+\delta\rho_{r,i}\right)=a_{i}^2\delta\rho_i\,,
\end{align}
where we imposed that initially $\Phi'_i=0$, we neglected gradient terms and we used the subscript “$i$” to denote evaluation at the initial time. In the Newton gauge, initial curvature fluctuations $\Phi_i$ are then proportional to $\delta\rho_i$. Thus, adiabatic initial conditions correspond to $S_i=0$ and $\Phi_i\propto\delta\rho_i\neq 0$ and isocurvature initial conditions to $\Phi_i\propto\delta\rho_i=0$ and $S_i\neq 0$. An interesting perspective from these definitions is that adiabatic initial conditions are set by the fluid with dominant energy density (since $S=0$ the dominant $\rho$ has also dominant $\delta\rho$)  while isocurvature initial conditions are mainly given by the sub-dominant field (since $\delta\rho_m+\delta\rho_r=0$ and $S$ depends inversely in $\rho_m$).

The closed system of equations for curvature-isocurvature fluctuations in Fourier modes is given by (see App.~\ref{app:cosmoformulas} or \cite{Domenech:2020ssp,Domenech:2021and} for details)
\begin{align}\label{eq:eomPhi}
    \Phi''+3{\cal H}(1+c_s^2)\Phi'+({\cal H}^2(1+3c_s^2)+2{\cal H}')\Phi+c_s^2k^2\Phi=\frac{a^2}{2}\rho_mc_s^2S\,,
\end{align}
and
\begin{align}\label{eq:eomS}
    S''&+ 3{\cal H}c_s^2S'-\frac{3}{2a^2\rho_r}c_s^2{k^4\Phi}+\frac{3\rho_m}{4\rho_r}c_s^2k^2 S=0\,,
\end{align}
where we defined as usual
\begin{align}
c_s^2\equiv\frac{4}{9}\frac{\rho_r}{\rho_m+4\rho_r/3}\,.
\end{align}
The relative velocity $V_{\rm rel}$ can be computed from $S$ by
\begin{align}\label{eq:vrelwiths}
V_{\rm rel}=S'/k^2\,.
\end{align}
Let us note here that in the matter dominated universe where $c_s^2\to 0$ the general “growing mode” solution to Eqs.~\eqref{eq:eomPhi} is $\Phi={\rm constant}$. The precise value of the constant will be set by the evolution during the radiation dominated universe. We present first the general solutions for superhorizons scales and later the solutions for general $k$ in the radiation dominated universe. We will denote with $S_i$ and $\Phi_i$ the initial values of isocurvature and curvature in the far past, formally when $a\to 0$. The conclusions do not change if $a$ has a non-zero value as the solutions are attractors \cite{Lozanov:2023aez}.\\

\paragraph{\textbf{Superhorizon fluctuations} ($k\ll {\cal H}$):} If we drop all terms containing $k$ in Eqs.~\eqref{eq:eomPhi} and \eqref{eq:eomS}, and re-write them using $a$ as a time variable, we find that the general solutions are given by \cite{Kodama:1986ud,mukhanov2005physical,Durrer:2004fx}
\begin{align}
S(\xi)&=S_i\,,\\
\Phi(\xi)&=\Phi_i \left(\frac{8}{5\xi^3}\left(
\sqrt{1+\xi}-1\right)
-\frac{4}{5\xi^2}
+\frac{1}{5\xi}+\frac{9}{10}\right)\nonumber\\&
+S_i\left(\frac{16}{5\xi^3}
   \left(1-\sqrt{1+\xi}\right)+\frac{8}{5 \xi^2}
   -\frac{2}{5\xi}+\frac{1}{5}\right)\,,
\end{align}
where for compactness we defined $\xi=a/a_{\rm eq}$. We see that while isocurvature remains constant on superhorizon scales, curvature fluctuations $\Phi$ are not. For the adiabatic component we have that
\begin{align}\label{eq:phisuperad}
\Phi_{\rm ad}(a)\approx\Phi_i\times\left\{
\begin{aligned}
&1 &(a\ll a_{\rm eq})\\
&\frac{9}{10} &(a\gg a_{\rm eq})
\end{aligned}
\right.\,,
\end{align}
where the factor $9/10$ actually comes from the conservation on $\zeta$ and the relation $\zeta=\tfrac{5+3w}{3+3w}\Phi$ for constant $w=p/\rho$. For isocurvature fluctuations we instead have
\begin{align}\label{eq:phisuperiso}
\Phi_{\rm iso}(a)\approx S_i\times\left\{
\begin{aligned}
&\frac{1}{8}\frac{a}{a_{\rm eq}} &(a\ll a_{\rm eq})\\
&\frac{1}{5} &(a\gg a_{\rm eq})
\end{aligned}
\right.\,.
\end{align}
Thus, while we have vanishing initial curvature perturbation, i.e. $\Phi_{\rm iso}(a\to0)\to 0$, it grows as ${a}/a_{\rm eq}$ and saturates to $1/5$ of the initial isocurvature after matter-radiation equality.\\

\paragraph{\textbf{Fluctuations during radiation domination} ($a\ll a_{\rm eq}$):} To study the evolution of general fluctuations during radiation domination, we shall take the limit $\tau\ll \tau_{\rm eq}$ ($a\ll a_{\rm eq}) $in Eqs.~\eqref{eq:eomPhi} and\eqref{eq:eomS}. At leading order, they are given by \cite{Domenech:2021and}
\begin{align}\label{eq:eomPhiRD}
    \frac{d^2\Phi}{dx^2}+\frac{4}{x}\frac{d\Phi}{dx}+\frac{1}{3}\Phi+\frac{1}{4\sqrt{2}\kappa x}\left(x\frac{d\Phi}{dx}+(1-x^2)\Phi-2S\right)\simeq 0\,,
    \end{align}
and
    \begin{align}\label{eq:eomSRD}
    \frac{d^2S}{dx^2}&+\frac{1}{x}\frac{dS}{dx}-\frac{x^2}{6}\Phi-\frac{1}{2\sqrt{2}\kappa}\left(\frac{dS}{dx}-\frac{x}{2}S-\frac{x^3}{12}\Phi\right)\simeq 0\,,
    \end{align}
where we defined for compactness
    \begin{align}
    x=k\tau \quad {\rm and}\quad\kappa=\frac{k}{\keq}\,.
    \end{align}
In these new variables, the limit of interest is given by $x\ll \kappa$ (or $k_{\rm eq}\tau\ll 1$). Due to the length of the solutions we treat the initial adiabatic and initial isocurvature cases separately below.

For initial curvature fluctuations, $\Phi_i\neq 0$ and $S_i=0$, we solve the leading order terms in Eqs.~\eqref{eq:eomPhiRD} and \eqref{eq:eomSRD}, namely we first solve the homogeneous equation for $\Phi$ and plug it in in the equation for $S$. This yields
\begin{align}\label{eq:analyticalsolad}
    \Phi_{\rm ad}(x/\kappa \ll 1)&\approx 3{\Phi_i} \frac{j_1(c_s x)}{c_s x}+\mathcal{O}\left(x/\kappa\right)\,,\\
\label{eq:analyticalsolSad}
    S_{\rm ad}(x/\kappa \ll 1) &\approx 9\Phi_i\left(\gamma_E-\frac{1}{2}+\frac{1}{2}\cos (c_sx)-{\rm Ci}(c_sx)+\log(c_sx)\right)+\mathcal{O}\left(x/\kappa\right)\,,
\end{align}
where $\gamma_E\approx 0.577$ and ${\rm Ci}(x)$ is the cosine integral function. Note that 
Eq.~\eqref{eq:analyticalsolad} is the standard solution for adiabatic perturbations in the radiation universe. $\Phi$ is first constant and decays as $x^{-2}$ once a given mode enters the sound horizon, that corresponds to $c_sx>1$. We also see that $S$ is negligible for $c_sx<1$ as it is proportional to $x^4$ but grows as $\log(c_s x)$ for $c_sx>1$. Such logarithmic grows is due to the fact that matter perturbations grow logarithmically during the radiation dominated universe \citep{Voruz:2013vqa}. It should be noted that one may also use Eq.~\eqref{eq:analyticalsolSad} to compute the next order solution to $\Phi$. We refer the interested reader to Ref.~\cite{Kodama:1986ud} for general solutions.

For the initial isocruvature case, $\Phi_i=0$ and $S_i\neq0$, we follow Ref.~\cite{Domenech:2021and}. In this case, we see that we can expand the solution of $\Phi$ and $S$ in powers of $\kappa^{-1}$ for $\kappa\gg1$ and in powers of $x/\kappa$ for $\kappa\ll1$, as e.g. $S=S_i+S_1+...$ and $\Phi=\Phi_1+...$ etcetera. The leading contribution to Eq.~\eqref{eq:eomPhiRD} $\Phi$ is a then a constant $S$ and we may also compute the effects of the leading solution to $\Phi$ to the next leading solution to $S$. Doing so we find \cite{Domenech:2021and}
\begin{align}\label{eq:analyticalsol}
    \Phi_{\rm iso}(x/\kappa \ll 1)&\approx \frac{3S_i}{2\sqrt{2}\kappa} \frac{1}{x^3} \left(6+x^2-2\sqrt{3}x\sin(c_sx)-6\cos(c_sx)\right)+\mathcal{O}\left(x/\kappa\right)^{2}\,,\\
\label{eq:analyticalsolS}
    S_{\rm iso}(x/\kappa \ll 1) &\approx S_i+\frac{3S_i}{2\sqrt{2}\kappa}\left(x+\sqrt{3}\sin(c_sx)-2\sqrt{3}{\rm Si}(c_sx)\right)+\mathcal{O}\left(x/\kappa\right)^{2}\,,
\end{align}
where ${\rm Si}(x)$ is the sine integral function. Looking at $x\ll 1$ we see that initially the curvature perturbation grows as $\Phi_{\rm iso}\propto x$, reaches a maximum at around $c_sx\sim 1$ and then decays as $x^{-2}$. It is also interesting to note that for $c_sx\gg1$ we have $\Phi_{\rm ad}\supset \sin(c_sx)$ and $\Phi_{\rm iso}\supset \cos(c_sx)$, recovering the well-known result that adiabatic and isocurvature initial conditions give an out of phase curvature fluctuations.
Isocurvature $S$ is constant for $c_sx<1$ and then grows with $x$ for $c_sx>1$. Interestingly, it is possible that $S$ reaches a high enough amplitude for PBHs to form \cite{Passaglia:2021jla}. Although we will not explore this possibility in this work, let us write down the time when the local density of matter is larger than that of radiation. This happens at
\begin{align}\label{eq:xnl}
    \tau_{\rm NL}=\frac{\sqrt{2}}{\keq S_i}\,.
\end{align}
At times $\tau>\tau_{\rm NL}$ we can no longer trust of linear solutions. Nevertheless, this only occurs when $S_i$ is very large. In most situations regarding induced GWs we may consider that $\tau_{\rm NL}$ is late enough such that it does not affect the production of GWs as curvature perturbations already decayed significantly. For example, requiring that the non-linear regime occurs deep inside the horizon (i.e. $x_{\rm NL}>1$) translates into an upper bound on the initial isocurvature, namely
\begin{align}\label{eq:sibound}
    S_i<\sqrt{2}\kappa \,,
\end{align}
for a given isocurvature mode with wavenumber $k$.\\

\begin{figure}
\includegraphics[width=0.49\columnwidth]{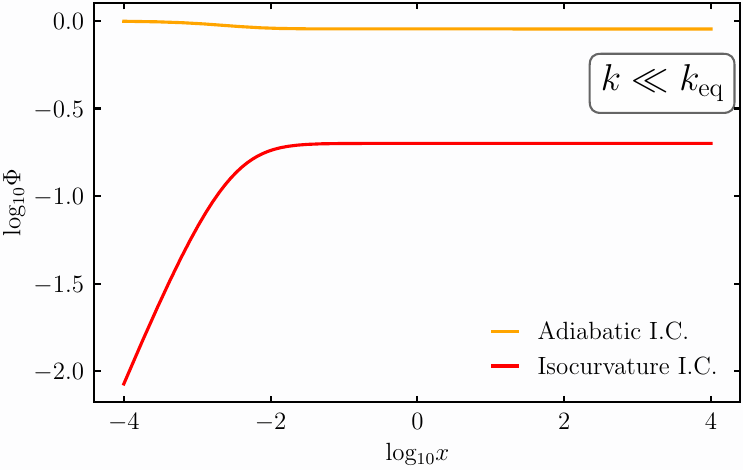}
\includegraphics[width=0.49\columnwidth]{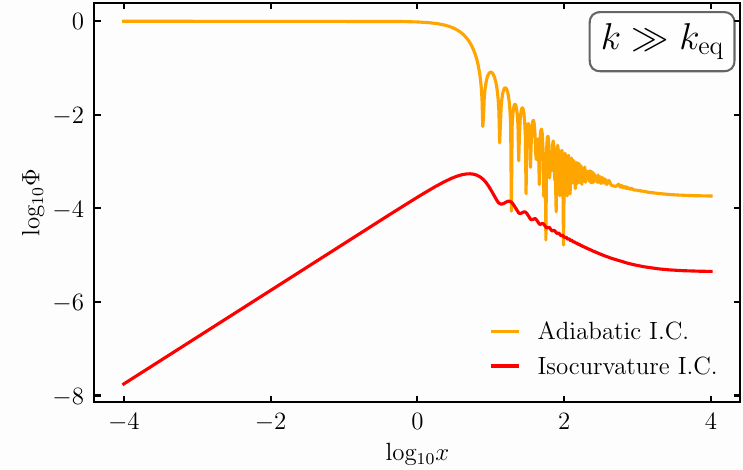}
\caption{Numerical solutions to the evolution of curvature fluctuations for adiabatic and isocurvature initial conditions in terms of $x=k\tau$, respectively in purple and orange lines. We normalized the amplitude of the initial conditions to $\Phi_i=1$ and $S_i=1$. On the left we show the behavior on scales with $k\ll k_{\rm eq}$. For the numerical solution we chose $k=10^{-3}k_{\rm eq}$. See how both lines saturate after matter-radiation equality to a constant. For the adiabatic case the constant is $9/10$ and for isocurvature $1/5$. On the right we instead show $k\gg k_{\rm eq}$, in particular $k=500k_{\rm eq}$. In this case the lines also saturate to a constant after matter-radiation equality. However, for $k\gg k_{\rm eq}$ the amplitude after $\tau_{\rm eq}$ is of the order of $(k_{\rm eq}/k)^2$ in both cases. \label{fig:phisub}}
\end{figure}

\paragraph{\textbf{Fluctuations during matter domination} ($a\gg a_{\rm eq}$):}  Well inside matter domination curvature fluctuations on all scales becomes constant. The amplitude of such fluctuations is then determined by whether they were superhorizon or subhorizon during the radiation domination epoch. For $k\ll k_{\rm eq}$ its value is given by Eqs.~\eqref{eq:phisuperad} and \eqref{eq:phisuperiso} respectively for adiabatic and isocurvature initial conditions. For $k\gg k_{\rm eq}$ the amplitude has a suppression factor proportional to $(k/k_{\rm eq})^{-2}$, which comes from the $x^{-2}$ decay during the radiation dominated phase. For the analytical approximations, we refer the reader to the work of Kodama and Sasaki \cite{Kodama:1986ud}, although in our simplified set up the numerical prefactors could be refined by matching at matter-radiation equality. As it is not crucial for our purposes we leave it as an exercise. Most important for the present review is the result of the curvature perturbation for isocurvature initial conditions at matter domination, which is given by \cite{Kodama:1986ud}
\begin{align}\label{eq:phisuperisomatter}
\Phi_{\rm iso}(a\gg a_{\rm eq})\approx S_i\times\left\{
\begin{aligned}
&\frac{1}{5} &(k\ll k_{\rm eq})\\
&\frac{3}{4}\left(\frac{k}{k_{\rm eq}}\right)^{-2} &(k\gg k_{\rm eq})
\end{aligned}
\right.\,.
\end{align}
We show the results of numerical integration in Fig.~\ref{fig:phisub}. In the numerical results we find that the coefficient for $k\gg k_{\rm eq}$ in Eq.~\eqref{eq:phisuperisomatter} is close to $1$. For easier comparison with the literature though, we maintain the coefficient of Eq.~\eqref{eq:phisuperisomatter} as it only introduces small errors.

\subsection{General formulation of isocurvature induced gravitational waves}

We proceed with the formal solution to induced gravitational waves. Our starting point is the equations of motion for secondary GWs \eqref{eq:linearGmunu2} which is given by
\begin{align}\label{eq:linearGmunu3}
h_{ij}''+2{\cal H}h'_{ij}+\Delta h_{ij}={\cal P}_{ij}\,^{ab}S_{ab}\,,
\end{align}
where $S_{ij}$ \eqref{eq:source2} using the linear Einstein Equations is given by
\begin{align}
S_{ij}&=4\partial_i\Phi\partial_j\Phi+6c_s^2\frac{\rho}{\rho_r}\partial_i\left(\frac{\Phi'}{{\cal H}}
+\Phi\right)\partial_j\left(\frac{\Phi'}{{\cal H}}+\Phi\right)+6a^2c_s^2\rho_m\partial_iV_{\rm rel}\partial_jV_{\rm rel}\,.
\end{align}
It will be more convenient to work in Fourier modes, where we give a primordial initial amplitude of $S_k(0)$ and/or $\Phi_k(0)$ and describe the evolution of a given mode by a transfer function, say $T_\Phi(k\tau)$. In the notation of the previous subsections $S_k(0)$ should be understood as $S_i$ for a given mode $k$. The same applies to $\Phi$. We will also only consider the case of either adiabatic or isocurvature initial condition and, therefore, we neglect any cross term coming from the gradient squared terms. Doing so Eq.~\eqref{eq:linearGmunu3} reads
\begin{equation}\label{eq:eominducedGW3}
h''_{\mathbf{k},\lambda}+2{\cal H} h'_{\mathbf{k},\lambda}+k^2 h_{\mathbf{k},\lambda}={\cal S}_{\mathbf{k},\lambda}\,,
\end{equation}
with
\begin{equation}
{\cal S}_{\mathbf{k},\lambda}=4\int \frac{d^3q}{(2\pi)^3} e_\lambda^{ij}(k)q_iq_j \left\{
\begin{aligned}
\Phi_{\mathbf{q}}(0)\Phi_{|\mathbf{k-q}|}(0)\\
S_{\mathbf{q}}(0)S_{|\mathbf{k-q}|}(0) 
\end{aligned}
\right\}
f(\tau, q,|\mathbf{k-q}|)\,,
\end{equation}
where $\Phi_{\mathbf{q}}(0)$ and $S_{\mathbf{q}}(0)$ respectively refer to initial adiabatic or initial isocurvature fluctuations, and we defined
\begin{align}\label{eq:f}
    f(\tau, q,|\mathbf{k-q}|)=&f(x,k, u,v)=T_\Phi(v x)T_\Phi(u x)+ \frac{3}{2}c_s^2 a^2\rho_mT_{V_{\rm rel}}(v x )T_{V_{\rm rel}}(u x)\nonumber\\&+\frac{3}{2}c_s^2\left(1+\frac{\rho_m}{\rho_r}\right)\left(T_\Phi(v x)+\frac{T'_\Phi(v x)}{\cal H}\right)\left(T_\Phi(u x)+\frac{T'_\Phi(u x)}{\cal H}\right)\,.
\end{align}
In Eq.~\eqref{eq:f} we introduced for later use the notation
\begin{align}
v=q/k\quad,\quad u=|\mathbf{k-q}|/k\,.
\end{align}
Also, in Eq.~\eqref{eq:f}, $T_{V_{\rm rel}}(x)$ is the transfer function of $V_{\rm rel}$ which is not important in the regimes of interest, so we will not consider it in the following sections. Note that the transfer functions that appear in Eq.~\eqref{eq:f} have to be properly chosen according to whether we are considering adiabatic or isocurvature initial conditions.

Assuming that the isocurvature fluctuations are Gaussian we arrive at a compact expression for the spectral density of induced GWs \cite{Espinosa_2018,Kohri:2018awv,Domenech:2021and}, namely
    \begin{align}\label{eq:Phgaussian}
    \Omega_{\rm GW,c}(k)=\frac{2}{3}\int_0^\infty dv\int_{|1-v|}^{1+v}du\left(\frac{4v^2-(1-u^2+v^2)^2}{4uv}\right)^2\overline{I^2(x_c,k,u,v)}{{\cal P}_{S}(ku)}{{\cal P}_{S}(kv)}\,,
    \end{align}
where ${\cal P}_{S}(k)$ is the initial dimensionless spectrum of isocurvature fluctuations, defined by
    \begin{align}
    \langle S_\mathbf{k}(0)S_{\mathbf{k}'}(0)\rangle=\frac{2\pi^2}{k^3}{\cal P}_S(k)\times(2\pi)^3\delta^{(3)}\left(\mathbf{k}+\mathbf{k}'\right)\,.
    \end{align}
For adiabatic initial conditions ${\cal P}_{S}(k)$ should be replaced by ${\cal P}_{\Phi}(k)$ at the initial time. In Eq.~\eqref{eq:Phgaussian} the notation $x_c$ refers to a time when the GW is well inside the horizon such that the spectral density is constant \cite{Inomata:2016rbd}. In Eq.~\eqref{eq:Phgaussian} we introduced the kernel which incorporates the information from the transfer functions and it is given by
\begin{equation}\label{eq:kernel1}
    I(x,k,u,v)\equiv x \,\int_{x_i}^{x}d\tilde x  \,G(x,\tilde x)f(\tilde x,k,u,v)\,,
\end{equation}
where $G(x,\tilde x)$ is the tensor mode’s Green’s function.

As we will be mostly interested in the induced GWs generated during an radiation dominated phase, either the one preceding matter domination or the one after an early matter domination, we restrict ourselves to the radiation dominated universe. In particular, the Green‘s function in the radiation dominated universe reads
\begin{align}\label{eq:greensG}
    G(x,\tilde x)= \frac{a(\tilde x)}{a( x )}\left(\sin x \cos \tilde x-\cos x \sin \tilde x \right)\,.
\end{align}
Then, we may split the kernel \eqref{eq:kernel1} into a sine and cosine terms as
\begin{align}\label{eq:transferI}
    I(x,k,u,v)=I_c(x,k,u,v)\sin x-I_s(x,k,u,v)\cos x\,,
\end{align}
where
\begin{align}\label{eq:Ics}
    I_{c/s}(x,k,u,v)\equiv\int_{0}^{x}d\tilde x  \, \tilde x\,\left\{
    \begin{aligned}
    &\cos \tilde x\\
    &\sin \tilde x
    \end{aligned}
    \right\} f(\tilde x,k,u,v)\,.
\end{align}
This allows us to take the oscillation average of the spectral density, which is given by
\begin{equation}\label{eq:kernel2}
    \overline{I^2(x\to\infty, k, u, v)} \simeq \frac{1}{2}\left(I_{c,\infty}^2(k,u,v)+I_{s,\infty}^2(k,u,v)\right)\,,
\end{equation}
where we took the limit $x\to\infty$ as we are interested in GW frequencies which are well within the horizon. It is interesting to note that at this point the main difference between the isocurvature and adiabatic initial condition cases is the different behavior of the transfer functions. The integrals $I_s$ and $I_c$ \eqref{eq:Ics} can be analytically carried out as they are integrals of trigonometric functions. However, the expressions are considerably long and so we will only present them in simplified situations. In the next sections we show in more detail the different GW spectrum from isocurvature and adiabatic initial conditions.

Before moving on, we write down the explicit relation between $\Omega_{\rm GW, c}$ \eqref{eq:Phgaussian} and the spectral density of GWs measured today. A straightforward relation can be found in the case when the radiation domination era where $\Omega_{\rm GW, c}$ is computed corresponds to the standard radiation dominated stage, i.e. prior to BBN and CDM domination. In that case, taking into account the redshifting of the GW energy density one finds that \cite{Domenech:2021ztg}
\begin{align}
\Omega_{\rm GW,0}h^2\approx 1.62\times 10^{-5}\,\Omega_{\rm GW, c}(k)\,,
\end{align}
where $\Omega_{\rm GW,0}$ is the energy density fraction of GWs evaluated today and $h=H_0/(100 \rm km/s/Mpc)$. $H_0$ is the Hubble parameter evaluated today. For simplicity, we assumed the standard model of particle physics. In the general case there is a dependence on the effective degrees of freedom, see e.g. \cite{Inomata:2016rbd,Saikawa:2018rcs}.

\section{ Gravitational waves from CDM isocurvature \label{sec:GWCDM}}

Let us consider that the matter component with isocurvature fluctuations is the CDM. This means that the initial fraction of CDM $\beta$ is fixed its the current abundance, which according to Planck \cite{Planck:2018vyg} is $\Omega_{\rm CDM,0}h^2\approx 0.12$. For our practical purposes though, we just need to use that the time of matter-radiation equality, normalized to today, is at $a^{-1}_{\rm eq}\approx 3400$ and that the comoving size of the universe at that time was $k_{\rm eq}\approx 0.01 \,{\rm Mpc}^{-1}$. To grasp the magnitude suppression factor $k_{\rm eq}/k$, let us write down the GW frequency in terms the relevant wavenumbers, that is
\begin{align}
f_{\rm GW}\sim 2\times \frac{k}{2\pi}\approx 3\,{\rm Hz}\left(\frac{k}{10^{15}\,{\rm Mpc}^{-1}}\right)\,,
\end{align}
where the first factor $2$ comes from the fact that two scalar modes source one induced tensor mode.
Thus, for the frequencies of interest, say between $\rm nHz$ and $\rm kHz$, we have suppression factors respectively between $10^{-8}$ and $10^{-19}$. This means that initial isocurvature has to be roughly of the inverse order of magnitude so that isocurvature induced GWs are detectable. It should be noted, though, that such large values of initial isocurvature are compatible with cosmological perturbation theory up to the time $\tau_{\rm NL}$. For a detailed discussion on the validity of perturbations see the appendix of Ref.~\cite{Domenech:2021and}. The current challenge is then not the validity of perturbations but to find a model with such large initial isocurvature. At the end of this section we will discuss some interesting cases where isocurvature need not be that large.

\begin{figure}
\includegraphics[width=0.55\columnwidth]{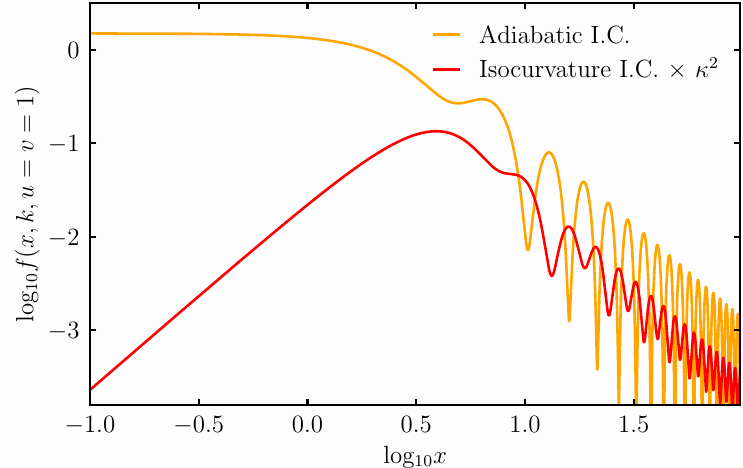}
\caption{Source term \eqref{eq:f} for induced GWs for adiabatic and isocurvature initial conditions, respectively in purple and orange, in terms of $x=k\tau$. For simplicity we chose $u=v=1$ but the main behavior is independent on such particular values. Here we consider $\tau\ll\tau_{\rm eq}$ and, therefore, the source term only decays after a given mode enters the horizon during radiation domination. See how for adiabatic initial conditions the source term is initial constant and then decays with large amplitude oscillations. Instead for isocurvature initial conditions the source term initially grows and then decays with smaller amplitude oscillations. Note that the frequency of the oscillations is the same in both cases but are out of phase. \label{fig:sourcecomparison}}
\end{figure}

Regarding isocurvature induced GWs, we are mostly interest in the small scale power spectrum. This constitutes fluctuations on scales which entered the horizon much before matter-radiation equality. Therefore, it is a very good approximation to use our solution Eq.~\eqref{eq:analyticalsol} during the radiation dominated epoch.  Plugging in Eq.~\eqref{eq:analyticalsol} into the integrals \eqref{eq:Ics} we find that
\begin{align}
    \label{eq:Icinf}
    I_{c,\infty}(k,u,v)&=\frac{9}{32u^4v^4\kappa^{2}}\Bigg\{-3u^2v^2+\left(-3+u^2\right)\left(-3+u^2+2v^2\right)\ln\left|1-\frac{u^2}{{3}}\right|\nonumber\\&
    +\left(-3+v^2\right)\left(-3+v^2+2u^2\right)\ln\left|1-\frac{v^2}{{3}}\right|\nonumber\\&
    -\frac{1}{2}\left(-3+v^2+u^2\right)^2\ln\left[\left|1-\frac{(u+v)^2}{{3}}\right|\left|1-\frac{(u-v)^2}{{3}}\right|\right]\Bigg\}\,,
    \end{align}
    and
    \begin{align}
    \label{eq:Isinf}
    I_{s,\infty}(k,u,v)&=\frac{9\pi}{32u^4v^4\kappa^{2}}\Bigg\{9-6v^2-6u^2+2u^2v^2+\left(3-u^2\right)\left(-3+u^2+2v^2\right)\Theta\left(1-\frac{u}{\sqrt{3}}\right)\nonumber\\&
    +\left(3-v^2\right)\left(-3+v^2+2u^2\right)\Theta\left(1-\frac{v}{\sqrt{3}}\right)\nonumber\\&
    +\frac{1}{2}\left(-3+v^2+u^2\right)^2\left[\Theta\left(1-\frac{u+v}{\sqrt{3}}\right)+\Theta\left(1+\frac{u-v}{\sqrt{3}}\right)\right]\Bigg\}\,,
    \end{align}
where we took the limit $x\to \infty$ for analytical simplicity. With the kernels Eqs.~\eqref{eq:Icinf} and \eqref{eq:Isinf} we are ready to compute the isocurvature induced GW spectrum via Eq.~\eqref{eq:Phgaussian}. Let us emphasize that Eqs.~\eqref{eq:Icinf} and \eqref{eq:Isinf} are valid for any primordial spectrum of isocurvature fluctuations. The only requirement is that the relevant fluctuations enter the horizon much before matter-radiation equality. The averaged kernel for GWs induced by adiabatic fluctuations can be found in App.~\ref{app:adiabaticiGWs}. For illustration, we also compare the source term \eqref{eq:f} between initially adiabatic and isocurvature fluctuations in Fig.~\ref{fig:sourcecomparison}.

It is interesting to note that the explicit $\kappa$ dependence in the kernel \eqref{eq:Icinf} and \eqref{eq:Isinf} can be absorbed via the definition of an effective power spectrum of curvature perturbations, namely
\begin{align}\label{eq:Pphieffectve}
    {\cal P}_\Phi^{\rm eff}(k)\equiv\kappa^{-2}{\cal P}_{\cal S}(k)\,.
\end{align} 
Although this is a mere redefinition, it illustrates the differences with the adiabatic initial conditions. For instance, after using ${\cal P}_\Phi^{\rm eff}(k)$ the averaged kernel resembles that of the adiabatic initial conditions case, in the sense that they have the same resonance and same scaling in the limits of integration \cite{Domenech:2021and}.  The detailed shape is of course different. Also note that the GW spectrum has a global suppression proportional to $\kappa^{-4}$, consistent with our expectations that $\Phi\propto \rho_m/\rho_r \times S\propto \kappa^{-1} S$ for isocurvature initial conditions when evaluated at horizon crossing. We proceed to discuss some concrete examples.

\subsection{GWs induced by a peaked primordial CDM isocurvature spectrum}

To have an idea of the shape of the isocurvature induced GW spectrum and its difference with the standard (curvature) induced GWs, let us consider that the primordial spectrum of CDM isocurvature fluctuations peaks at given scale, say $k_p$. For concreteness and to follow previous works we consider a log-normal peak given by
\begin{align}\label{eq:lognormalpeak}
\mathcal{P}_{S}(k)=\frac{\mathcal{A_S}}{\sqrt{2\pi}\Delta}\exp\left[-\frac{\ln^2(k/k_p)}{2\Delta^2}\right],
\end{align}
where $A_S$ is the amplitude and $\Delta$ is the logarithmic width of the peak. This is standard practice when dealing with GWs induced by primordial adiabatic fluctuations (see e.g. \cite{Pi:2020otn}) and it is well motivated from inflationary models \cite{Kawasaki:1997ju,Frampton:2010sw,Kawasaki:2012wr,Inomata:2017okj,Pi:2017gih,Cai:2018tuh,Cai:2019jah,Chen:2019zza,Ashoorioon:2019xqc,Chen:2020uhe,Garcia-Bellido:1996mdl,Yokoyama:1998pt,Kohri:2012yw,Clesse:2015wea,Cheng:2016qzb,Espinosa:2017sgp,Inomata:2017okj,Kannike:2017bxn,Garcia-Bellido:2017mdw,Ando:2017veq,Cheng:2018yyr,Ando:2018nge,Espinosa:2018eve,Inomata:2018cht,Braglia:2020eai,Palma:2020ejf,Fumagalli:2020adf}. Similar conclusions extend to the isocurvature case \cite{Pi:2021dft,Han:2022kvj}. We note that the formal limit of $\Delta\to 0$ corresponds to a Dirac delta peak, namely
\begin{align}
\mathcal{P}_{S}(k)=\mathcal{A_S}\delta(\ln(k/k_p))\,.
\end{align}
A log-normal is considered to be “sharp” if $\Delta\lesssim 0.1$ and broad otherwise \cite{Pi:2020otn}. We take a similar ansatz for primordial curvature fluctuations $\Phi_i$.

The Dirac delta case is convenient as it allows for an analytical expression for $\Omega_{\rm GW,c}$ simply by replacing $u=v=k_p/k$ in Eqs.~\eqref{eq:Icinf}, \eqref{eq:Isinf} and \eqref{eq:Phgaussian}. Explicitly, we obtain
\begin{align}\label{eq:omegacdirac}
    \Omega_{\rm GW,c}(k)&= \frac{{\cal A}_S^2}{3} \left(\frac{k}{k_p}\right)^{-2} \left(1-\frac{k^2}{4k_p^2}\right)^2  \left(I^2_{c,\infty}\left(\frac{k_p}{k},\frac{k_p}{k}\right) + I^2_{s,\infty}\left(\frac{k_p}{k},\frac{k_p}{k}\right)\right)   \Theta\left(2 k_p-k\right),
\end{align}
where the sharp cut-off at $k=2k_p$ comes from momentum conservation. From Eq.~\eqref{eq:omegacdirac} we see that, as in the initially adiabatic case (see e.g. the discussions in Ref.~\cite{Domenech:2021ztg}), the induced GW spectrum has a resonant peak at $k=2c_sk_p$ and decays as $k^2\ln^2k$ in the low frequency tail, namely for $k\ll k_p$  \cite{Cai:2018dig,Yuan:2019wwo,Cai:2019cdl}. However, contrary to the adiabatic case, there is no destructive interference with vanishing GW spectrum at $k=\sqrt{2}c_sk_p$. This is because adiabatic initial conditions lead to coherent, order unity, oscillations of $\Phi$ while for isocurvature initial conditions the oscillations of $\Phi$ are modulations with a different phase. Thus, one can in principle distinguish the adiabatic and isocurvature cases by the shape of the GW spectrum around $k\sim k_p$. One of the important differences, though, is that the isocurvature induced GW spectrum has a suppressed amplitude, namely the peak amplitude is given by
\begin{align}\label{eq:omegacdiracpeak}
    \Omega^{\rm iso,peak}_{\rm GW,c}&\propto {{\cal A}_S^2}\times \left(\frac{k_{\rm eq}}{k_p}\right)^4\,.
\end{align}
This means that the amplitude of the initial power spectrum of isocurvature fluctuations must be very large to partly compensate for the suppression, roughly ${{\cal A}_S}\propto \left({k_{p}}/{k_{\rm eq}}\right)^2\gg1$. Such large values of initial isocurvature fluctuations are not in contradiction with the validity of cosmological perturbation. The reason is that any effect of isocurvature is accompanied by a factor $\rho_m/\rho_r$ during the radiation dominated era. The product of initial isocurvature times $\rho_m/\rho_r$ is always smaller than unity during the relevant times for our calculations.

\begin{figure}
\includegraphics[width=0.49\columnwidth]{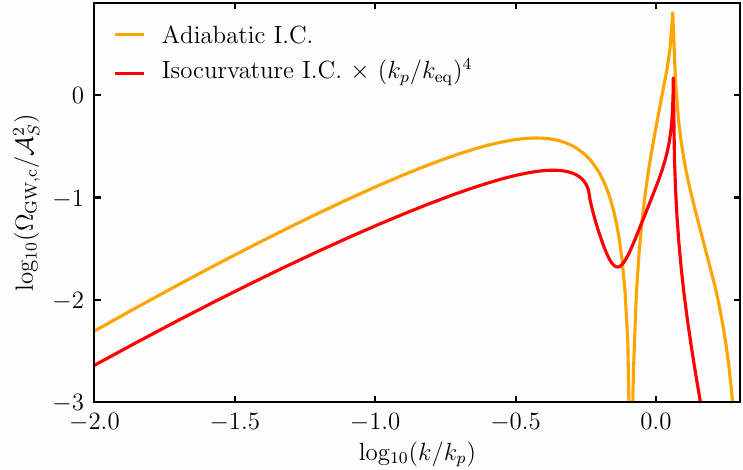}
\includegraphics[width=0.49\columnwidth]{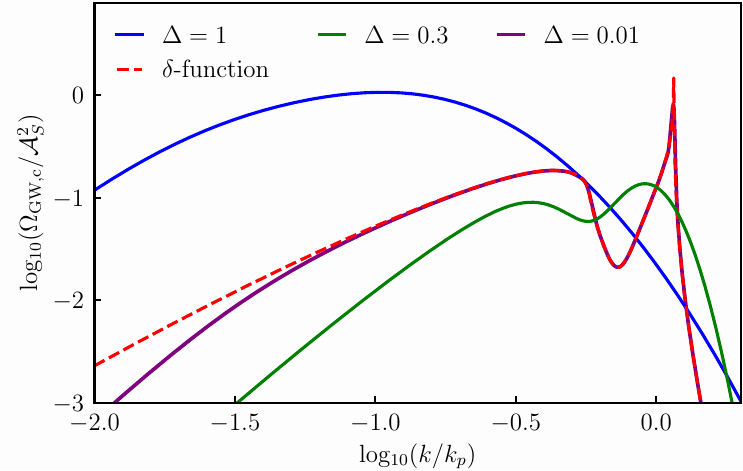}
\caption{Induced GW spectrum from a log-normal primordial spectrum in terms of comoving wavenumber $k$ normalized to the peak wavenumber $k_p$. The amplitude of the GW spectrum is normalized such that ${\cal A}_{S}=1$ and ${\cal A}_{\Phi}=1$ in Eq.~\eqref{eq:lognormalpeak} for $\mathcal{P}_{S}$ as well as $\mathcal{P}_{\Phi}$, i.e. primordial isocurvature and curvature respectively. On the left we respectively show the induced GWs from primordial curvature and primordial isocurvature in purple and orange, in the case of a Dirac delta power spectrum. See how the overall shape is similar, that is a resonant peak at $k=2c_sk_p$ and a $k^2$ low frequency tail with a logarithmic running. However, the shape around the peak and the dip is different which breaks the degeneracy of the GW signals. On the right we show the isocurvature induced GW spectrum from a log-normal peak with $\Delta=0.01,0.3,1$ respectively in red, green and blue. The orange dotted line show the Dirac delta result of the left figure. See how the peak of the induced GW spectrum shifts to lower $k$ for larger $\Delta$ as explained by Eq.~\eqref{eq:Pphieffectve}. \label{fig:isoinducedGW}}
\end{figure}

For the log-normal peak we recover similar results as the Dirac delta for $\Delta\lesssim 0.01$, except that for scales $k<\Delta\, k_p$, where the slope transitions to a $k^3$ scaling, as in the adiabatic case \cite{Pi:2020otn}. For $\Delta\gtrsim 0.01$ similar conclusions apply but because of the $\kappa^{-2}$ dependence in ${\cal P}_\Phi^{\rm eff}$ \eqref{eq:Pphieffectve}, coming from the fact that modes that enter earlier or more suppressed, the peak of the GW spectrum moves to lower values of $k$. We show the numerical results for the Dirac delta and log-normal peak in Fig.~\ref{fig:isoinducedGW}.

\subsection{Remarks, issues and future work}

So far we have discussed that a large amplitude of primordial isocurvature fluctuations is compatible with perturbation theory, as long as they satisfy Eq.~\eqref{eq:sibound}. We also assumed that the initial isocurvature fluctuations can be considered as Gaussian. Such assumption, while expected to give a rough approximation, is strictly speaking not correct. The reason is that isocurvature fluctuations are mostly density fluctuations of the matter fluid, i.e. $S\propto \delta\rho_m/\rho_m$, and while ${\cal P}_S\gg1$ is certainly possible, the probability distribution of $S$ cannot be Gaussian as $S>-1$ (i.e. $\rho_m+\delta\rho_m>0$). Thus, a correct distribution function would be highly skewed, starting at $S=-1$ and with large tails for large S, so as to keep the average $\langle S \rangle=0$. The main issue here is that in the absence of a concrete realization it is difficult to parametrize the distribution function as it cannot be expanded in terms of Gaussian distributions. Some estimates are given in the appendix of Ref.~\cite{Domenech:2021and} but the general expectation is that large tails lead to large 4-point functions and a larger amplitude of the induced GWs. Thus, what we computed in the previous section might well be a lower bound on the amplitude of CDM isocurvature induced GWs. It would be interesting to study a concrete case and confirm these expectations. 

\begin{figure}
\includegraphics[width=0.5\columnwidth]{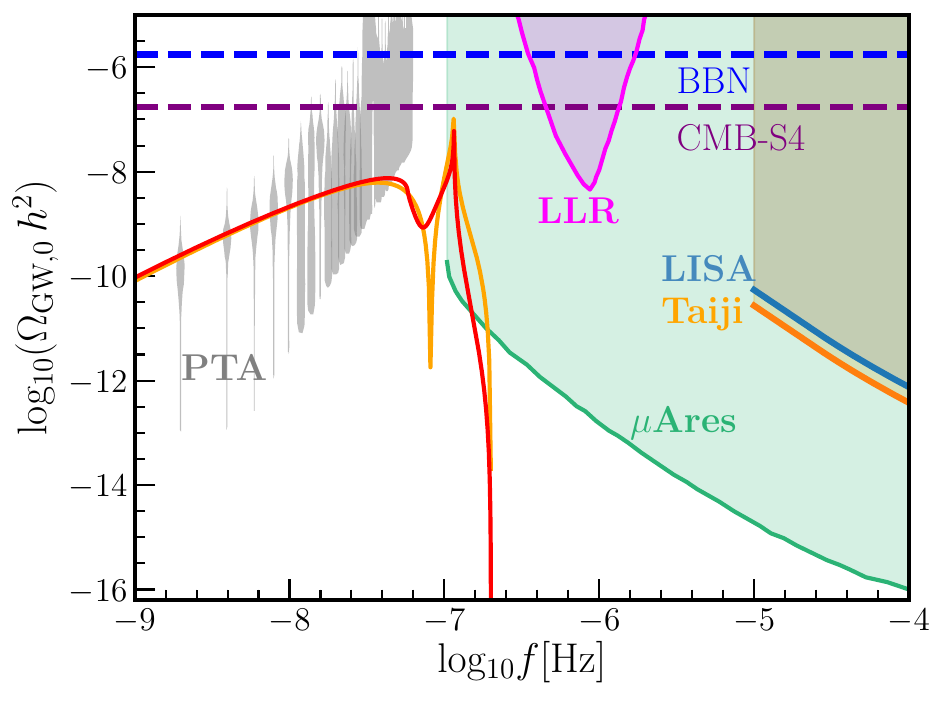}
\caption{Example of induced GWs from a Dirac delta spectrum for primordial curvature and isocurvature fluctuations. As an illustration we chose $A_\Phi=0.01$ and $A_S=0.05 \,(k_{p}/k_{\rm eq})^2\sim 10^{18}$ to fit the PTA data by eye. The peak is chosen at $f_p\sim 10^{-7}\,\rm Hz$ ($k_p\sim 10^7\,{\rm Mpc}^{-1}$). Gray violins indicate the recent NANOGrav results \cite{NG15-NP} and in blue and orange we show power-law integrated sensitivity curves~\cite{Thrane:2013oya} for LISA and Taiji \cite{Barke:2014lsa,Ruan:2018tsw}. See \cite{ce,A+,voyager,Schmitz:2020syl} for the sensitivity curves. The horizontal blue and purple dashed lines respectively show the current constraint from BBN \cite{Cyburt:2004yc,Arbey:2021ysg,2023arXiv230112299G} and future constraints from CMB-S4 experiments \cite{2016arXiv161002743A,Arbey:2021ysg}. We also include future sensitivity of $\mu$-Ares \cite{Sesana:2019vho} and Lunar Laser Ranging from Ref.~\cite{Blas:2021mqw} (see also \cite{Fedderke:2021kuy}). GW detectors such as $\mu$-Ares may tell apart whether the GWs were induced by adiabatic or isocurvature initial conditions. \label{fig:isoPTA}}
\end{figure}

It is also interesting to consider the hypothetical case where the possible GW background signal reported by the PTA collaborations  \cite{NG15-SGWB,NG15-pulsars,EPTA2-SGWB,EPTA2-pulsars,EPTA2-SMBHB-NP,PPTA3-SGWB,PPTA3-pulsars,PPTA3-SMBHB,CPTA-SGWB,InternationalPulsarTimingArray:2023mzf} is due to isocurvature induced GWs. Although we will not carry out a detailed data analysis, we may infer a good order of magnitude estimate for the required amplitude for the primordial spectrum of CDM isocurvature fluctuations. From the analysis of GWs induced by primordial adiabatic fluctuations we have that, if given by a Dirac delta spectrum, one needs ${\cal A}_\Phi\sim 10^{-2}-10^{-1}$ and a peak position around $f_p\sim 10^{-7}\,\rm Hz$ \cite{NG15-NP}. This means that for the primordial CDM isocurvature one requires ${\cal A}_S (k_{\rm eq}/k_p)^2\sim{\cal A}_\Phi$, which roughly corresponds to ${\cal A}_S\sim 10^{18}-10^{19}$. While this value of ${\cal A}_S$ is certainly big it is still within the validity range of our induced GW spectrum calculations as ${\cal A}_S (k_{\rm eq}/k_p)^2\ll 1$. With future detectors in the $\mu$Hz window, such as $\mu$-Ares \cite{Sesana:2019vho} it may be possible to see the peak of the GW spectrum and to distinguish the nature of the initial conditions. We present an example in Fig.~\eqref{fig:isoPTA}. 

\section{ Gravitational waves from the PBH dominated early universe \label{sec:PBHdom}}

An interesting model which includes an early stage of matter domination ended by a fast transition to radiation domination is the PBH reheating scenario \cite{Lidsey:2001nj,Hidalgo:2011fj}. In this scenario, tiny PBHs are formed after inflation with a fraction large enough such that they dominate the early universe before evaporating via Hawking radiation. The most interesting part is that, as shown by Refs.~\cite{Inomata:2019ivs,Inomata:2020lmk}, a sudden transition from matter to radiation domination greatly enhances the spectrum of induced GWs. In our case, this generates a distinct GW signal of a PBH dominated early universe.

\subsection{PBH formation and evaporation}

Consider that initially we have a universe filled with a homogeneous radiation fluid. Then, at some moment in time, PBHs form roughly at the same time and with the same mass $M_{\rm PBH}$ (this is often called a monochromatic mass function). We assume that PBHs formed by the collapse of large primordial fluctuations (although it does not have to be necessarily the case) and so their mass at formation is related to the Hubble parameter at the time of formation, say $H_{\rm f}$, via $M_{\rm PBH,f}\sim 4\pi M_{\rm pl}^2/H_{\rm f}$. Note that fact that PBHs formed from adiabatic primordial fluctuations does not affect our simplified picture of an initial homogeneous radiation fluid, as we shall see. We also assume from now on that the particle content of the universe after evaporation is given by the standard model of particles particle. Additional particles might change the precise coefficients through the effective degrees of freedom. We refer the interested reader to Refs.~\cite{Inomata:2020lmk,Domenech:2020ssp} for the details.

After formation, PBH make a fraction $\beta$ of the total energy density at formation. It is interesting to relate the $\beta$ to the number density of PBHs, namely
\begin{align}
  \beta=\frac{\rho_{\rm PBH,f}}{\rho_{r,\rm f}}\sim \frac{4\pi}{3H_{\rm f}^3}n_{\rm PBH,f}\,,
\end{align}
where we used that $\rho_{\rm PBH,f}=M_{\rm PBH,f}\times n_{\rm PBH,f}$ and $\rho_{r, \rm f}\approx 3H_{\rm f}^2M_{\rm pl}^2$. Thus, $\beta$ can be interpreted as the \textit{mean} number of PBHs for Hubble volume. We emphasize that it refers to the \textit{mean} number because in general there will be statistical number density fluctuations.

Now, while the number density of PBHs is conserved, which in an expanding universe means that $n_{\rm PBH}\propto a^{-3}$, the PBH mass decreases due to Hawking radiation. One finds that the mass decays as \cite{Inomata:2020lmk}
\begin{align}\label{eq:MPBH(t)}
M_{\rm PBH}(t)\approx M_{{\rm PBH},f}\left(1-\frac{t}{t_{\rm eva}}\right)^{1/3}\,,
\end{align}
where $t_{\rm eva}$ is the time of evaporation and is approximately given by
\begin{align}\label{eq:teva}
t_{\rm eva}\sim \frac{M_{\rm PBH,f}^3}{M_{\rm pl}^4}\sim 0.4 \,{\rm fs}\times \left(\frac{M_{\rm PBH,f}}{10^4\,{\rm g}}\right)^{3}\,,
\end{align}
where ${\rm fs}=10^{-15}\,{\rm s}$ are femtoseconds. Here $t$ is cosmic time related to conformal time by $dt=ad\tau$. Since we assumed that all PBHs have the same mass and that formed roughly at the same time, they all evaporate at the same time as well. From the Hubble parameter at the time of evaporation ($H_{\rm eva}\sim 1/t_{\rm eva}$) we can estimate the temperature of the radiation fluid filling the universe right after evaporation. In doing so, we find that
\begin{align}
T_{\rm eva}\approx 2.76\times 10^4\,{\rm GeV}\left(\frac{M_{{\rm PBH},f}}{10^4{\rm g}}\right)^{-3/2}\,.
\end{align}
From the Hubble parameter we can also compute the comoving size of the Hubble horizon at evaporation, namely \cite{Domenech:2020ssp}
\begin{align}\label{eq:keva}
k_{\rm eva}={\cal H}_{\rm eva}\approx 4.7\times 10^{11}{\rm Mpc}^{-1}\left(\frac{M_{{\rm PBH},f}}{10^4{\rm g}}\right)^{-3/2}\,.
\end{align}
As expected, the larger the PBH mass the later the reheating and the larger the comoving horizon at evaporation. It should be noted that PBHs must reheat the universe well before BBN, which imposes that $T_{\rm eva}>4\,{\rm MeV}$ and translates into $M_{{\rm PBH},f}<5\times 10^8\,{\rm g}$.  We can also compute the minimum value of $\beta$ such that PBHs eventually dominate the universe before evaporating, which is given by
\begin{align}\label{eq:betamin}
\beta>6\times10^{-10}\left(\frac{M_{{\rm PBH},f}}{10^4{\rm g}}\right)^{-1}\,.
\end{align}

\begin{figure}
\includegraphics[width=0.5\columnwidth]{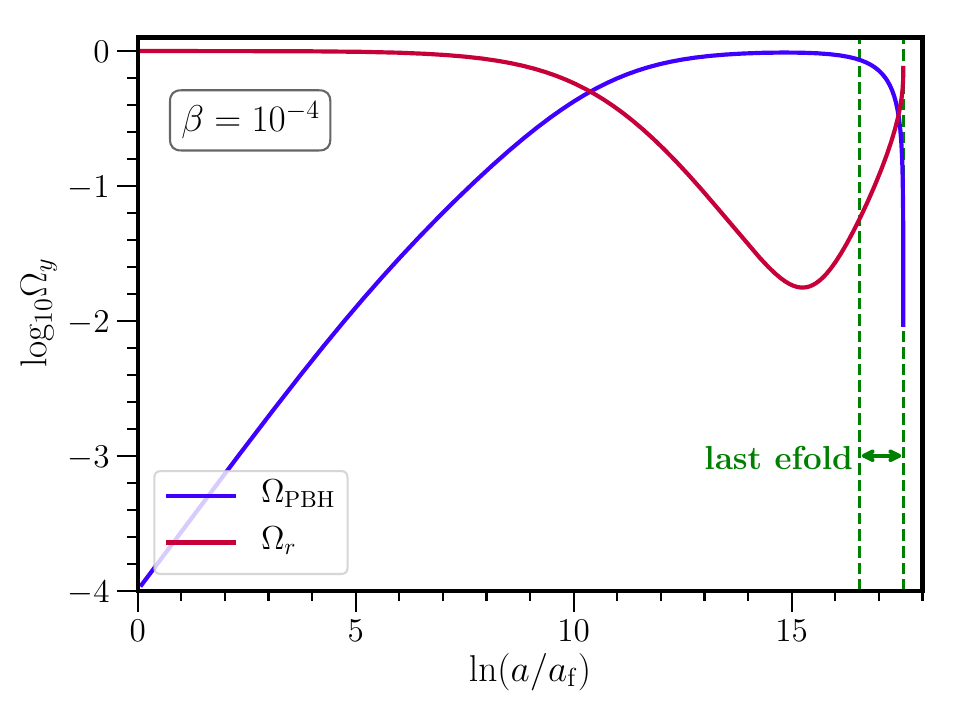}
\caption{Energy density fraction $\Omega_y=\rho_y/(3H^2M_{\rm pl}^2)$, with $y={m,r}$ respectively for PBH (matter) in blue and radiation in red, in terms of the scale factor normalized at formation. We chose the initial PBH fraction to be $\beta=10^{-4}$ for illustrative purposes. See how PBH domination can last for several e-folds and that evaporation is faster than one e-fold. \label{fig:PBHrho}}
\end{figure}

When dealing with fluctuations, we treat the collection of PBHs as an almost pressureless matter fluid. The evaporation is then considered as the decay of “matter” into radiation via a decay rate given by
\begin{align}\label{eq:Gamma}
\Gamma\equiv-\frac{d\ln M_{\rm PBH}}{dt}\,.
\end{align}
Then, the energy density of the PBH fluid is gradually transferred to radiation and obeys
\begin{align}
\dot\rho_{\rm PBH}+\left(3H+\Gamma\right)\rho_{\rm PBH}=0\,,
\end{align}
where $\dot\,\equiv d/dt$. The equation for the radiation energy density is similar but with the opposite sign in front of $\Gamma$, as to satisfy energy conservation. One can check that the total evaporation of PBHs is not instantaneous but takes about a quarter of an e-fold. We show in Fig.~\ref{fig:PBHrho} the evolution of $\rho_{\rm PBH}$ and $\rho_r$ for a particular example. 

The main point that we want to emphasize here is that fluctuations with $k\gg \Gamma$ have a typical time scale much larger than the evaporation rate and, as such, they are very much affected by the finite duration of evaporation \cite{Inomata:2020lmk}. This translates into a suppression factor for curvature fluctuations with $k\gg \Gamma$ after evaporation given by
\begin{align}\label{eq:xi}
{\cal S}_\Phi(k)\equiv \frac{\Phi_{\rm lRD}}{\Phi^{\rm instant}_{\rm lRD}}\approx\left(\frac{k}{k_{\rm eva}}\right)^{-1/3} \,,
\end{align}
where “lRD” means late radiation domination and “instant” refers to the amplitude of $\Phi$ obtained by an instant matching from matter to radiation domination. Note that the exponent $1/3$ in Eq.~\eqref{eq:xi} is directly related to the exponent in the decay of the PBH mass \eqref{eq:MPBH(t)}. The relation between $\Phi$ and $\rho_{\rm PBH}$ follows from the Poisson equation on very subhorizon scales, that is $k^2\Phi\sim a^2\rho_{\rm PBH}\delta_{\rm PBH}$. We must take the suppression factor ${\cal S}_\Phi(k)$ \eqref{eq:xi} into account when computing any transfer function for the curvature perturbation. We suggest to read Ref.~\cite{Inomata:2020lmk} for more details on the calculations.

\subsection{PBH number density fluctuations and initial isocurvature}

While on average PBH are homogeneously distributed, the fact that PBHs are discrete objects introduces inhomogeneities. For instance, the mean PBH separation is given by \cite{Papanikolaou:2020qtd}
\begin{align}
d_{\rm f}\equiv\left(\frac{3}{4\pi n_{{\rm PBH},{\rm f}}}\right)^{1/3}\approx\beta^{-1/3}{H_{\rm f}}^{-1}\,.
\end{align}
On scales larger than $d_{\rm f}$ we have a coarse grained fluid picture but on scales smaller than $d_{\rm f}$ we either see a PBH or not. Now, since PBH formation is a rare event (the number of PBH per Hubble volume at formation $\beta$ is very small), we can approximate the process of formation as PBHs appearing randomly and uniformly distributed in space (in other words, everywhere has the same probability of hosting a PBH). And, this means that the probability of having $n$ PBHs in a given volume is described by Poisson statistics. This allows us to compute the variance of number density fluctuations, which for Poisson fluctuations is constant in wavenumber and given by \cite{Papanikolaou:2020qtd}
\begin{align}\label{eq:deltanpbh}
\left\langle\frac{\delta n_{\rm PBH,f}(k)}{n_{\rm PBH,f}}\frac{\delta n_{\rm PBH,f}(k')}{n_{\rm PBH,f}}\right\rangle=\frac{4\pi}{3}\frac{d_{\rm f}^3}{a_{\rm f}^3}\,\delta(k+k')\,,
\end{align}
where the factor $a_{\rm f}$ comes from the fact that we are working with comoving scales.
In terms of dimensionless quantities, a constant physical variance implies a dimensionless variance proportional to $k^3$. Thus, the variance of fluctuations increases with wavenumber until the fluid picture is no longer valid, which occurs at an ‘‘ultra-violet’’ cut-off given by the mean inter-PBH comoving separation, namely
\begin{align}\label{eq:kuv}
k_{\rm UV}=a_{\rm f}/d_{\rm f}\approx\beta^{1/3}{{\cal H}_{\rm f}}=\beta^{1/3}{{k}_{\rm f}}\,,
\end{align}
where we used that $k_{\rm f}={\cal H}_{\rm f}$.

Such initial PBH number density fluctuations are in fact initial isocurvature fluctuations. Simply put, by energy conservation any missing part of the radiation fluid (that ended up in a PBH) is compensated by the PBHs themselves, such that at formation we have $\delta\rho_{\rm PBH,f}+\delta\rho_{r,\rm f}=0$. This implies that
\begin{align}\label{eq:isocurvature2}
S_f=\frac{\delta \rho_{\rm PBH,f}}{\rho_{\rm PBH,f}}-\frac{3}{4}\frac{\delta \rho_{r,\rm f}}{\rho_{r,\rm f}}\approx\frac{\delta \rho_{{\rm PBH},f}}{\rho_{{\rm PBH},f}}\approx\frac{\delta n_{{\rm PBH},f}}{n_{{\rm PBH},f}}\,,
\end{align}
where we used that initially $\rho_{\rm PBH}\ll \rho_r$. From Eq.~\eqref{eq:deltanpbh}, the dimensionless initial isocurvature power spectrum then reads
\begin{align}\label{eq:PS}
{\cal P}_{S}(k)=\frac{2}{3\pi}\left(\frac{k}{k_{\rm UV}}\right)^3\,.
\end{align}
This is the initial isocurvature that will eventually generate induced GWs \cite{Papanikolaou:2020qtd,Domenech:2020ssp}.

\subsection{GWs from PBH isocurvature fluctuations after PBH evaporation}

PBH number density fluctuations provide isocurvature initial conditions for cosmological fluctuations. However, as we discussed, induced GWs are mainly sourced by curvature fluctuations. Thus, before computing the induced GWs we have to compute the transfer functions for $\Phi$. As we will be interested in the GWs sourced by curvature fluctuations after evaporation and on very small scales (since $k_{\rm UV}\gg k_{\rm eva}$), we just have to follow $\Phi$ until the end of the PBH dominated era. There are of course GWs induced during the early radiation domination and the early matter domination phase. However, they turn out to be subdominant and, therefore, we neglect them. See Ref.~\cite{Papanikolaou:2020qtd} for the GWs induced during the PBH dominated phase.

To compute the transfer functions, we start noting that the comoving Hubble parameter at equality is proportional to $\beta$, that is
\begin{align}
{k_{\rm eq}}={\sqrt{2}\beta}\,k_{\rm f}\,,
\end{align}
where $k_{\rm eq}={\cal H}_{\rm eq}$. It then follows that we have the resulting hierarchy: $k_{\rm f}\gg k_{\rm UV}\gg k_{\rm eq}\gg k_{\rm eva}$. To put it into words, the initial fluctuations on scales similar to $k_{\rm UV}$ are initially superhorizon (they are on scales larger than $k_{\rm f}$). But, these scales around $k_{\rm UV}$ enter the horizon well before matter-radiation equality and so $k_{\rm UV}\gg k_{\rm eq}$. Then at evaporation all fluctuations of interest are largely subhorizon. This means that we can use the second line of the transfer function in Eq.~\eqref{eq:phisuperisomatter}, namely the transfer function for isocurvature initial conditions, together with the suppression factor \eqref{eq:xi}, which yields
\begin{align}
T_{\Phi}(t_{\rm eva}; k\gg k_{\rm eq})=T_{\Phi_{\rm iso}}(k){\cal S}_\Phi(k)=\frac{3}{4}\left(\frac{k}{k_{\rm eq}}\right)^{-2}\left(\frac{k}{k_{\rm eva}}\right)^{-1/3}\,.
\end{align}

To compute the GWs induced after evaporation we have to continue the constant value of $\Phi$ during the PBH (matter) domination to the late radiation domination. After matching, we find that 
\begin{align}\label{eq:phi}
\Phi_{\rm lRD}(k\tau)=\frac{1}{c_sk\bar\tau}\left(C_1j_1(c_sk\bar \tau)+C_2y_1(c_sk\bar \tau)\right)\quad;
\quad\bar\tau\equiv \tau-\tau_{\rm eva}/2,
\end{align}
where
\begin{equation}\label{eq:c1c2}
C_1=-\Phi_{\rm eva}(k)\left(c_sk\tau_{\rm eva}/{2}\right)^{2}\cos(c_sk\tau_{\rm eva}/2)\quad,
\quad
C_2=C_1 \tan(c_sk\tau_{\rm eva}/2)\,.
\end{equation}
Now, with the evolution of $\Phi$ we can re-compute the kernel for the induced GWs. Here we only provide the main steps of the calculations and we refer the reader to Ref.~\cite{Inomata:2020lmk,Domenech:2020ssp} for the details. 

The most important point is to realize that, while the amplitude of $\Phi$ in Eq.~\eqref{eq:phi} starts from a constant and quickly decays, the amplitude of its time derivative, $\Phi'$, begins at zero (by continuity) and suddenly jumps to an amplitude proportional to $k/k_{\rm eva}$, which is huge for $k\sim k_{\rm UV}$. The reason for this jump is that during the matter dominated era $\Phi$ remained constant but the relevant scales became more and more subhorizon. Then, at the late radiation domination, $\Phi$ resumes its decay but with very fast oscillations. Yet, since it has not decayed during the matter dominated phase, the amplitude of $\Phi'$ is suddenly very large by a factor $a_{\rm eva}/a_{\rm UV}\sim k_{\rm UV}/k_{\rm eva}$ (the supposed decay if there never were a matter dominated era). 

Another explanation for the enhancement is given in Ref.~\cite{Inomata:2019ivs}. During the matter dominated phase PBH density fluctuations grow, as standard CDM fluctuations do. But suddenly all those density fluctuations are converted into radiation which wants to propagate. The large density fluctuations become density waves with a huge velocity (note that by Einstein Equations in App.~\ref{app:cosmoformulas} $V\propto \Phi'/{\cal H}$). Inspecting the source term \eqref{eq:f} we see that the main contribution to the kernel \eqref{eq:kernel1} comes from $\Phi'$ and, basically, we can approximate the kernel by
\begin{align}\label{eq:kernelRD}
I_{\rm lRD}(x,u,v,x_{\rm eva})\approx\frac{\bar x}{2}uv\int_{x_{\rm eva}/2}^{\bar x} d{ {\bar x}_1}\,{ {\bar x}_1}^2\, G^{\rm lRD}(\bar x,{ {\bar x}_1})\frac{dT_\Phi^{\rm lRD}(u{ {\bar x}_1})}{d(u{ {\bar x}_1})}\frac{dT_\Phi^{\rm lRD}(v{ {\bar x}_1})}{d(v{ {\bar x}_1})}\,,
\end{align}
where $G^{\rm lRD}(\bar x,{ {\bar x}_1})$ is given by \eqref{eq:greensG}. After integration and selecting only those terms related to the resonant peak (those at $c_s(u+v)\sim 1 $ where the frequency of the source term is equal to that of the tensor mode) we arrive at
\begin{align}\label{eq:irdapp}
\overline{I^2_{\rm lRD,{\rm res}}}&(u+v\sim c_s^{-1},x_{\rm eva})\approx \frac{c_s^4u^2v^2}{2^{15}}x_{\rm eva}^8T_{\Phi}^2(t_{\rm eva}; vk)T_{\Phi}^2(t_{\rm eva}; uk){\rm Ci}^2(|1-(u+v)c_s|x_{\rm eva}/2)\,.
\end{align}

To compute the induced GW spectrum we plug in the averaged kernel \eqref{eq:irdapp} into Eq.~\eqref{eq:Phgaussian} and integrate. However, for sufficiently peaked spectrum the integral is well approximated by a power-law with a sharp cut-off \cite{Inomata:2020lmk,Domenech:2020ssp,Domenech:2021wkk}. A power spectrum is considered to be peaked enough if the integrand in Eq.~\eqref{eq:Phgaussian} grows for growing $u$ and $v$. Since it contains $(u v)^2\sim v^4$, the effective spectral index of $P_\Phi(k)$ should be larger that $-5/2$. Otherwise the integral has to be carried out numerically. Our effective power spectrum of curvature fluctuations at evaporation reads
\begin{align}
{\cal P}_\Phi^{\rm eva}(k\gg k_{\rm eq})\approx T^2_{\Phi_{\rm iso}}(k)\,{\cal S}^2_\Phi(k)\,{\cal P}_{S}(k)=\frac{3}{8\pi}\left(\frac{k}{k_{\rm eq}}\right)^{-4}\left(\frac{k}{k_{\rm eva}}\right)^{-2/3}\left(\frac{k}{k_{\rm UV}}\right)^{3}\,,
\end{align}
which has an effective spectral index of $-5/3$. We see that our effective power spectrum is well within the power-law approximation for the integral.

\begin{figure}
\includegraphics[width=0.49\columnwidth]{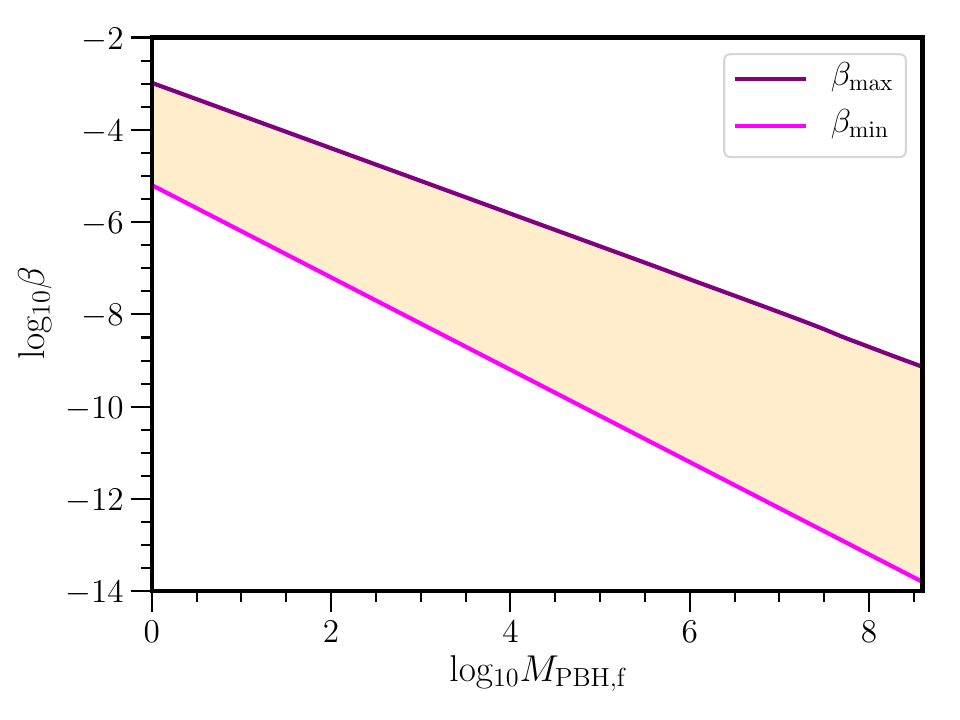}
\includegraphics[width=0.49\columnwidth]{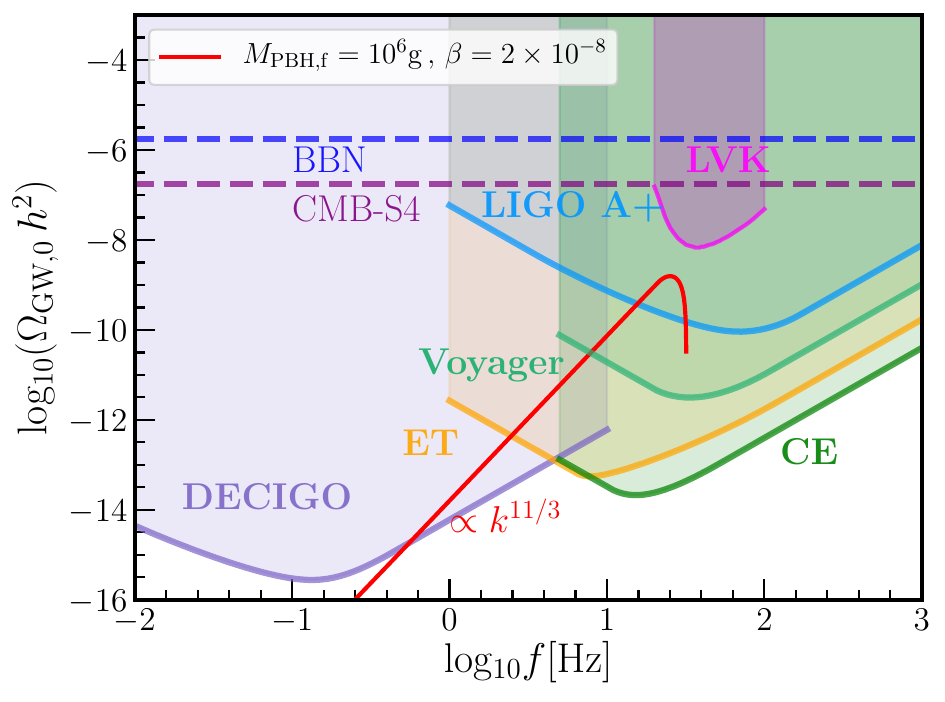}
\caption{On the left we show in the shaded orange region the parameter space for the initial fraction of PBHs, $\beta$, such that PBHs dominate the early universe ($\beta>\beta_{\rm min}$; magenta line) and that the induced GWs satisfies BBN bounds ($\beta<\beta_{\rm max}$; purple line), as a function of PBH mass. On the right we present an example of induced GWs from the PBH dominated early universe with $M_{\rm PBH,f}=10^6\,{\rm g}$ and $\beta=2\times 10^{-8}$. We also show the power-law integrated sensitivity curves~\cite{Thrane:2013oya} for DECIGO, Einstein Telescope (ET), Cosmic Explorer (CE), Voyager and LIGO A+ experiments \cite{ce,A+,voyager,Schmitz:2020syl} as well as the upper bounds from the LIGO/Virgo/KAGRA collaboration \cite{KAGRA:2021kbb}. The horizontal lines show BBN \cite{Cyburt:2004yc,Arbey:2021ysg,2023arXiv230112299G} and CMB-S4 experiments constraints, respectively in blue and purple \cite{2016arXiv161002743A,Arbey:2021ysg}. Finding a very peaked GW signal with a slope of $k^{11/3}$ above Hz frequencies might be an indication of the PBH reheating scenario.\label{fig:pbhdomsensititivty}}
\end{figure}

With the aforementioned approximations, the PBH isocurvature induced GW spectrum is roughly given by 
\begin{align}
\Omega_{\rm GW,c}\approx \Omega_{\rm GW,\rm res}\left(\frac{k}{k_{\rm UV}}\right)^{11/3}\Theta(k_{\rm UV}-k)\,,
\end{align}
where the peak amplitude of the GW spectrum is approximately given by
\begin{align}\label{eq:gwspeak}
\Omega_{\rm GW,\rm res}(k\sim k_{\rm UV})
&\approx \frac{1}{24576\pi\,2^{1/3}\sqrt{3}}\left(\frac{k_{\rm UV}}{k_{\rm eva}}\right)^{17/3} \left(\frac{k_{\rm eq}}{k_{\rm UV}}\right)^8
\approx 10^{30}\beta^{16/3}
\left(\frac{M_{{\rm PBH},f}}{10^4{\rm g}}\right)^{34/9}\,.
\end{align}
Furthermore, using Eqs.~\eqref{eq:kuv} and \eqref{eq:keva}, we find that the peak GW frequency is located at
\begin{align}
f_{\rm UV}
\approx 1700\, {\rm Hz}\left(\frac{M_{\rm PBH,f}}{10^4\,{\rm g}}\right)^{-5/6}\,.
\end{align}
Thus, on one hand, we see that for $5\times 10^8{\rm g}>M_{\rm PBH,f}>10^4\,{\rm g}$ the peak of the GW spectrum enters in the observable range of LIGO/VIRGO/KAGRA and future detectors such as ET and CE. It is interesting to note that the peak frequency only depends on the PBH mass at formation providing a clear probe of the PBH mass. On the other hand, the peak amplitude of the GW spectrum can be used to probe the initial PBH fraction. For instance, requiring that the peak amplitude is not in contradiction with current BBN constraints imposes that 
\begin{align}\label{eq:betamax}
\beta < 10^{-6}\left(\frac{M_{\rm PBH,f}}{10^4\,{\rm g}}\right)^{-17/24}\,.
\end{align}
This is the best constraint we have on the fraction of PBHs that evaporated before BBN. We show the resulting bounds on $\beta$ and an example of the GW spectrum from the PBH dominated universe in Fig.~\ref{fig:pbhdomsensititivty}.

\subsection{Remarks, issues and future work}

In our calculations we have mainly focus on the curvature perturbation $\Phi$ right after evaporation and we trusted our linear perturbation theory since $\Phi\ll 1$. However, when we look into the evolution of matter density fluctuations, which on small scales are related to $\Phi$ by the Poisson equation $2k^2\Phi=a^2\delta\rho$, we find that $\delta\rho/\rho$ at evaporation is larger than unity \cite{Papanikolaou:2020qtd,Domenech:2020ssp}. Depending on the PBH mass and fraction the amplitude of density fluctuations might be of a few orders of magnitude. This signals the on-set of non-linear physics, such as halo formation and PBH mergers. Unfortunately, one would need numerical simulations to explore such non-linear regimes. And, although it is an interesting question, it is out of the scope of this review. Note that we could also be conservative and stop our calculations of induced GWs when density fluctuations enter the non-linear regime. But, the fact is that we do not expect the GW production to stop by non-linearities. Thus, we take our result, Eq.~\eqref{eq:gwspeak}, as a rough order of magnitude estimate. For an interesting proposal in terms of turbulences in the radiation fluid after evaporation see Ref.~\cite{Kozaczuk:2021wcl}.

In this work we have also neglected the contribution for the adiabatic fluctuations that formed the PBHs, which for large PBH masses might enter the sensitivity of ET in the low frequency tail, and any primordial adiabatic fluctuations enhanced by the PBH dominated stage \cite{Inomata:2020lmk}. For works including both PBH isocurvature and adiabatic induced GWs see Refs.~\cite{Bhaumik:2020dor,Bhaumik:2022pil,Bhaumik:2022zdd}. For articles investigating the effect of a PBH dominated stage on the GW spectrum from cosmic strings see Refs.~\cite{Samanta:2021mdm,Borah:2022iym,Borah:2023iqo}. Other applications of the GWs from the PBH dominated universe can be found in Refs.~\cite{Papanikolaou:2021uhe,Papanikolaou:2022hkg,Banerjee:2022xft} in the context of modified gravity. Another interesting signal of the PBH dominated universe are GWs from Hawking evaporation itself, which could be important for highly spinning PBHs \cite{Domenech:2021wkk,Fujita:2014hha,Hooper:2020evu,Masina:2020xhk,Arbey:2021ysg,Kazemian:2022ihc,Ireland:2023avg}. For a similar application in the case where inflaton oscillons dominate the universe after inflation see Ref.~\cite{Lozanov:2022yoy}.

We end this section by emphasizing that the formalism employed is also applicable to any early matter dominated stage followed by a transition to radiation domination, as discussed in \cite{Domenech:2021wkk} (although see Ref.~\cite{Harigaya:2023ecg} for a transition to kinetic domination). For instance, if at the start of the early matter domination we have a spectrum of curvature fluctuations given by a power-law, for instance
\begin{align}\label{eq:pphidust}
{\cal P}_{\Phi}={\cal A}_\Phi\left(\frac{k}{k_{\rm UV}}\right)^{-n}\Theta(k_{\rm UV}-k)\,,
\end{align}
with some arbitrary cut-off at $k_{\rm UV}$, then the resulting induced GW spectrum after evaporation reads \cite{Domenech:2021wkk}
\begin{align}\label{eq:Phgaussiandust4}
\Omega_{\rm GWs,rh}(k\sim k_{\rm UV})\approx\Omega^{\rm peak}_{\rm GWs}\left(\frac{k}{k_{\rm UV}}\right)^{7-2n}\Theta(k_{\rm UV}-k)\,,
\end{align}
where
\begin{align}\label{eq:Phgaussiandust3}
\Omega^{\rm peak}_{\rm GWs}\approx\frac{\pi}{3\times 2^{12}c_s}\left(1-c_s^2\right)^2\left({4c^2_s}\right)^{n}\left(\frac{k_{\rm UV}}{k_{\rm rh}}\right)^7{\cal A}^2_{\Phi}\,.
\end{align}
It is also important to note that if transition is rather gradual, the enhancement of the induced GW spectrum disappears \cite{Inomata:2019zqy,Papanikolaou:2022chm}. This is particularly important if PBHs have a relatively broad mass function \cite{Inomata:2020lmk}.

\section{Discussions and conclusions \label{sec:conclusions}}

Initial isocurvature fluctuations may source induced GWs, in addition to any initial adiabatic fluctuations. One possibility is that isocurvature fluctuations do so via an induced curvature fluctuation, which is for example the case of initial CDM isocurvature. The revelant GWs are then induced during the standard radiation dominated era. In that situation, the amplitude of the induced GWs is roughly suppressed by a factor $\rho_m/\rho_r$ evaluated at horizon crossing for a given $k$ mode. This results in a large amplitude of initial isocurvature fluctuations, if such GWs enter the observable window of current and future GW detectors. Although a large amplitude  ($S\gg1$) is consistent with cosmological perturbation theory, it has the caveat that isocurvature fluctuations must be extremely non-Gaussian \cite{Domenech:2021and}. We expect that such non-Gaussianities may in fact enhance the production of induced GWs. As an example, we considered in Fig.~\ref{fig:isoPTA} the possibility that CDM isocurvature fluctuations explain the tentative GW signal reported by PTAs.

Another possibility is that the field mainly responsible for initial isocurvature fluctuations dominates the very early universe. Initial isocurvature fluctuations are then converted into curvature fluctuations which directly source induced GWs. As an example of this situation we considered the PBH dominated universe and the number density fluctuations coming from their Poisson statistics. When the PBH mass function is monochromatic, the final PBH evaporation is almost instantaneous, resulting in a large production of induced GWs right after evaporation. One may use this signal to probe the PBH reheating scenario and place constraints on the initial fraction of PBHs, as we showed in Fig.~\ref{fig:pbhdomsensititivty}. Although our estimate of the induced GW spectrum \eqref{eq:gwspeak} might be susceptible to non-linearities that occur during the PBH dominated phase and to the assumption of a monochromatic mass function, it provides hopes that we may be able to probe the PBH reheating scenario. 

As an extension of our result, we may wonder what implications would Planck remnants \cite{MacGibbon:1987my} at the end of evaporation have for the induced GW signal. See refs.~\cite{Chen:2014jwq,Hossenfelder:2012jw,Vidotto:2018hww,Eichhorn:2022bgu,Platania:2023srt} for reviews and models on black hole remnants. Recently in Ref.~\cite{Domenech:2023mqk}, it has been shown that if such remnants constitute the totality of the CDM then the initial PBH mass must be $5\times 10^5{\rm g}$ and the induced GW spectrum must peak at around $100\,{\rm Hz}$. Although it is a speculative possibility, such rather precise prediction for the peak frequency of the induced GWs might be one of the unique ways to test the PBH remnant scenario as CDM.

We would like to end this review with one interesting consequence of the formalism for early isocurvature induced GWs presented in this review. As argued in Refs.~\cite{Lozanov:2023aez,Lozanov:2023knf}, the formation of compact solitonic structures, such as oscillons, monopoles, domain walls, Q-balls, cosmic strings, etcetera, leads to a Poissonian distribution on scales much larger than the mean inter-soliton separation, just in the PBH scenario of \S~\ref{sec:PBHdom}. This means that statistical fluctuations of the number density of compact structures will lead to a dimensionless power spectrum proportional to $k^3$. Such power spectrum of isocurvature fluctuations first induces GW during the radiation dominated phase, as in \S~\ref{sec:GWCDM}, and might give another contribution if solitons dominate the early universe before decaying. Such generic prediction has been named “Universal Gravitational Waves of Solitons” \cite{Lozanov:2023aez,Lozanov:2023knf}.

We conclude by emphasizing that, while the consequences of primordial adiabatic fluctuations have been extensively studied, initial isocurvature fluctuations present new opportunities to test the physics of the unexplored very early universe.

\section*{Acknowledgments} 

I thank T.~Papanikolaou for useful correspondence and M.~Sasaki for useful discussions.  This research is supported by the DFG under the Emmy-Noether program grant no. DO 2574/1-1, project number 496592360. 

\appendix

\section{Equations for cosmological perturbations \label{app:cosmoformulas}}
For completeness we write in this appendix the main formulas for cosmological perturbation theory in the Newton gauge. The background equations are given by the Friedmann equations, which read
\begin{align}
&3M_{\rm pl}^2{\cal H}^2=a^2(\rho_m+\rho_r)\,,\\
&2M_{\rm pl}^2({\cal H}^2-2{\cal H}')=a^2(\rho_m+\rho_r+p_r)\,.
\end{align}

For the matter perturbations at linear level we have that $u_{m0}=u_{r0}=-a\left(1+\Psi\right),~u_{mi}=a\partial_iv_m$ and $u_{ri}=a\partial_iv_r$. We also perturb $\rho\to \rho+\delta\rho$. The trace-less part of the liner $ij$ Einstein equation yields
\begin{align}
\Phi+\Psi=0\,.
\end{align}
For simplicity we use $\Psi=-\Phi$ in all forthcoming equations. The $00$, $0i$ and $ij$ trace components lead us to
\begin{align}
&6{\cal H}\Phi'+6{\cal H}^2\Phi-2\Delta\Phi=a^2(\delta\rho_m+\delta\rho_r)\equiv a^2\delta\rho\,,\\
&\Phi'+{\cal H}\Phi=\frac{1}{2}a^2\left(\rho_mv_m+\frac{4}{3}\rho_rv_r\right)\equiv\frac{1}{2}a^2\rho V\,,\\
&\Phi''+3{\cal H}\Phi'+\left({\cal H}^2+2{\cal H}'\right)\Phi=-\frac{1}{6}a^2\delta \rho_r\,.
\end{align}
Energy conservation of the perfect fluid yields
\begin{align}
&\delta\rho_m'+3{\cal H}\delta\rho_m+\rho_m(3\Phi'+\Delta v_m)=0\,,\\
&\delta\rho_r'+4{\cal H}\delta\rho_r+\frac{4}{3}\rho_r(3\Phi'+\Delta v_r)=0\,.
\end{align}
Euler equations, or momentum conservation, gives
\begin{align}
&v_m'+{\cal H}v_m-\Phi=0\,,\\
&\rho_rv_r'+\frac{1}{4}\delta \rho_r-\rho_r\Phi=0\,.
\end{align}
The second order equations for tensor modes are given in the main text, Eq.~\eqref{eq:source}.

\section{Formula for induced GWs from adiabatic fluctuations\label{app:adiabaticiGWs}}
We take the analytical expression of GWs induced by adiabatic fluctuations from Ref.~\cite{Domenech:2021ztg}. In this case, the induced GW spectral density is given by
\begin{equation}\label{eq:Phgaussianfinal}
\Omega_{\rm GW,c}=\int_0^\infty dv\int_{|1-v|}^{1+v}du\,{\cal T}(u,v,c_s){{\cal P}_{\cal R}(ku)}{{\cal P}_{\cal R}(kv)}\,,
\end{equation}
where we defined for convenience
\begin{align}\label{eq:TRDPhgaussian}
    {\cal T}_{RD}(u,v,c_s)=\frac{2}{3}\left(\frac{4v^2-(1-u^2+v^2)^2}{4uv}\right)^2\overline{I^2(x_c,k,u,v)}\,,
\end{align}
and
\begin{align}\label{eq:w13}
{\cal T}_{RD}(u,v,c_s)=&\frac{y^2}{3c_s^4}\left(\frac{4v^2-(1-u^2+v^2)^2}{4u^2v^2}\right)^2\nonumber\\&\times
\left\{\frac{\pi^2}{4}y^2\Theta[c_s(u+v)-1]
+\left(1-\frac{1}{2}y \ln\left|\frac{1+y}{1-y}\right|\right)^2\right\}\,.
\end{align}
Lastly, we introduced the variable $y$ as it simplifies considerably the expressions, which reads
\begin{equation}\label{eq:y}
y=1-\frac{1-c_s^2(u-v)^2}{2c_s^2 uv}\,.
\end{equation}
The relation between $\Phi$ and ${\cal R}$ is given by
\begin{equation}\label{eq:pphitopR}
{\cal P}_{\Phi}=\frac{4}{9}{\cal P}_{\cal R}\,.
\end{equation}
This recovers the well-known result of Kohri and Terada \cite{Kohri:2018awv}.

\bibliography{refgwscalar.bib} 

\end{document}